\documentclass[11pt,a4paper]{article}
\usepackage{jstyle}
\usepackage{amsthm,bm,mathrsfs}
\usepackage{youngtab,simplewick}

\newcommand{\kp}[2]{\langle #1,#2\rangle}
\newcommand{\bp}[2]{[ #1,#2 ]}
\newcommand{\hk}{h_{\textrm{\tiny K}}}
\newcommand{\hj}{h_{\textrm{\tiny J}}}
\DeclareMathOperator\iso{\mathfrak{iso}}
\DeclareMathOperator\su{\mathfrak{su}}
\DeclareMathOperator\spl{\mathfrak{sl}}
\DeclareMathOperator\gl{\mathfrak{gl}}
\DeclareMathOperator\uu{\mathfrak{u}}
\DeclareMathOperator\syp{\mathfrak{sp}}

\author{Eduardo CONDE \quad}
\author{Euihun JOUNG \quad}
\author{Karapet MKRTCHYAN}

\affiliation{School of Physics \& Astronomy and Center for Theoretical Physics \\ Seoul National University, Seoul 08826 \rm KOREA}
\affiliation{Gauge, Gravity \& Strings, \ Center for Theoretical Physics of the Universe\\
Institute for Basic Sciences, Daejeon 34047 \rm KOREA}

\emailAdd{econdepe@snu.ac.kr}
\emailAdd{euihun.joung@snu.ac.kr}
\emailAdd{karapet@snu.ac.kr}

\title{\centering 
 Spinor-Helicity Three-Point Amplitudes\\
 from Local Cubic Interactions}
\abstract{We make an explicit link between 
the cubic interactions of off-shell fields  and the 
on-shell three-point amplitudes in four dimensions.
Both the cubic interactions and the on-shell three-point amplitudes
had been independently classified in the literature, but
their  relation has not been made explicit.
The aim of this note is to provide such a relation and 
discuss  similarities and differences of their constructions.
For the completeness of our analysis,
we also derive the covariant form of
all parity-odd massless vertices.
}

\begin{document}

\maketitle

\section{Introduction}

From the dawning of Quantum Field Theory, the relation between local fields
and unitary irreducible representations (UIRs) of Poincar\'e algebra has provided important guidelines in constructing 
field theory Lagrangians as well as in understanding various physical consequences of them.
For instance,  the relation between free fields and the corresponding UIRs
is very well understood by now:
one can either begin with the field equations to show that their general solutions
are in one-to-one correspondence with the UIRs (and their quantization leads to the Fock multi-particle space which corresponds to the tensor product representation of the UIR),
or reciprocally start from the UIRs and move to the Fock space, then finally construct the 
quantum fields as the operators transforming in a covariant fashion under Poincar\'e symmetries.
Concerning the familiar lower spin cases, 
one can find the detailed account \textit{e.g.} in \cite{Weinberg:1995mt}.

For these lower spins, the relation between free fields and UIRs can be further extended to the interacting level
in the sense that all consistent cubic interaction vertices 
allowed in a local field theory correspond to tri-linear invariant forms of the UIRs. 
Since the UIRs of Poincar\'e algebra --- typically labeled by the mass and spin $(m,s)$ 
and the spatial momenta and helicity state $(\vec p, h)$ ---
can be directly interpreted as the physical states, the tri-linear invariant forms
admit the interpretation of three-point amplitudes. 
From the representation point of view, the Clebsch-Gordan coefficients
must enjoy the same tri-linear invariance condition that the three-point amplitudes
satisfy. Hence these two objects actually coincide up to overall constants depending on $(m,s)$'s.
The latter constants remain arbitrary for the amplitudes unless the underlying theory
is fixed (the non-vanishing ones correspond, up to linear combinations, to the coupling constants of the theory),
but are fixed for the Clebsch-Gordan coefficients by
the completeness condition. 

This understanding in lower spins was soon extended towards more general UIRs.
The local free field theories for massive and massless higher-spin representations
were constructed in \cite{Singh:1974qz,Singh:1974rc} and \cite{Fronsdal:1978rb,Fang:1978wz}.
About the interactions, on the one hand there have been extensive studies about the Clebsch-Gordan
coefficients for the generic UIRs of Poincar\'e algebra (see \textit{e.g.} \cite{Strathdee:1967bi}).
On the other hand, from the field-theory point of view, all consistent
local cubic interactions of massless fields in four dimensions
have been derived in the light-cone gauge  \cite{Bengtsson:1983pd,Bengtsson:1986kh} then generalized to higher dimensions \cite{Fradkin:1991iy}. See \textit{e.g.} \cite{Berends:1984rq} for general discussions of this program.
However, 
due to various no-go results on the flat-space massless interactions \cite{Weinberg:1964ew,Coleman:1967ad,Aragone:1979hx} (see also \cite{Metsaev:1991mt,Metsaev:1991nb}) 
and the success of higher-spin theories in AdS
background \cite{Fradkin:1986qy,Vasiliev:1990en},
this direction of research lost its dynamics for a while.
A renewed interest on this issue came from, at least, two different directions.

The first direction is the AdS/CFT duality, which generically involves higher-spin fields on the AdS side
(even massless ones in certain cases, typically when the corresponding CFT becomes free).
Vasiliev's equations \cite{Vasiliev:1990en,Vasiliev:2001wa} describe the dynamics of massless fields
interacting with each other within the framework of the so-called unfolded formulation.
Even if the latter provides a fully consistent picture, it is still interesting
and illuminating to understand the duality  from a more mundane field-theoretical point of view.
Therefore, the interest on the nature of the flat-space cubic interactions was revived:
the light-cone vertices for massive and fermionic
fields were obtained in \cite{Metsaev:2005ar,Metsaev:2007rn},
and
the covariant form of the light-cone vertices was identified 
first for certain examples, \textit{e.g.}  \cite{Bekaert:2005jf,Boulanger:2006gr,Fotopoulos:2007yq,Zinoviev:2008ck,Manvelyan:2009vy,Manvelyan:2010wp}, then generalized in \cite{Manvelyan:2010jr,Sagnotti:2010at,Fotopoulos:2010ay,
Manvelyan:2010je,Metsaev:2012uy} to arbitrary spins.
For an overview of this line of investigations and an exhaustive list of references, the reader may
consult the review \cite{Bekaert:2010hw}.

The second direction is the ongoing progress in calculating scattering amplitudes 
using various on-shell methods (see \textit{e.g.} \cite{Elvang:2013cua,Dixon:2013uaa} and references therein). An important ingredient in exposing the simplicity
of certain four-dimensional scattering amplitudes (especially those
involving massless particles) is the use of spinor-helicity variables, as typically
exemplified by the Parke-Taylor $n$-gluon amplitude~\cite{Parke:1986gb,Berends:1987me}. Although {the helicity-spinor} formalism
was developed in the 80's mainly for phenomenological purposes, it possesses certain theoretical
advantages like making covariant properties manifest; and sparked by Witten's twistor
string~\cite{Witten:2003aa}, many theorists have adopted it in their works. As a byproduct of this
wave of activity, all possible structures of three-point amplitudes of massless particles were
identified in~\cite{Benincasa:2007xk} using spinor-helicity variables. Recently, in~\cite{Conde:2016vxs} the {classification} has been
extended to the cases involving massive particles.

The methods typically used in the higher-spin and amplitude communities do not share
the same philosophy: whereas Local Field Theory plays a prominent role
in the former, the latter tries to escape from it as much as possible.
It is therefore interesting to compare both methodologies and see
what are the points of agreement and disagreement, if any at all.
In this paper, we aim to make an explicit link between the  developments
in the two fields.

We consider both massless and massive particles, and 
show how the local Lagrangian vertices of Field Theory 
give rise to the known three-point amplitudes.
In doing so, instead of using the original light-cone form of the vertices,
we use their covariant version together with
several generalized Kronecker-delta identities,
valid only in four dimensions, to select the non-trivial vertices. 
For the completeness of massless interactions, we also rederive all parity-odd 
vertices in this way.
Regarding massive interactions, although our procedure is completely general,
we focus just on two types of interactions:
the first type involving only one massive particle and two massless ones
and the other type involving one massless particle and two massive particles of equal mass.

Besides providing an explicit link between the two results,
we hope that our analysis helps to understand better the interplay between
local Field Theory and the corresponding representation theory. 
Moreover, this work may also be considered as a toy exercise 
of the AdS/CFT duality where the local fields in flat space mimic
those in AdS while the three-point amplitudes 
play the role of the three-point correlation functions of the CFT. 
In fact, there have been several attempts to get the flat-space S-matrix
from AdS/CFT by taking a proper flat limit
(see for instance \cite{Penedones:2010ue,Fitzpatrick:2011ia} for a very general prescription).  

The organization of the paper is as follows.
In Section~\ref{sec: UIR}, we review the spinor realization of 
the massless and massive UIRs of the Poincar\'e algebra
with some discussions on its generality.
In Section~\ref{sec: massless}, we consider the massless case:
beginning with the local cubic vertices, we explicitly calculate
the corresponding three-point amplitudes using the spinor-helicity variables.
In particular, we show how the inclusion of both parity-even and -odd vertices 
exhaust all possible structures found from the amplitude side, up to
some subtleties.
In Section~\ref{sec: massive}, we move to the massive case:
after reviewing the massive cubic vertices of different types,
we focus on two cases. The case of one massive and two massless fields
is analyzed in Section~\ref{ssec: 1m}, while Section~\ref{ssec: 2m} contains the analysis 
of the case with two equal-mass and one massless particle.
In all these cases, we find a good agreement with
the result obtained from the amplitude side.
After presenting our conclusions in Section~\ref{sec:conc}, we include in
Appendix~\ref{app} some technical details regarding the massive UIRs of the Poincar\'e algebra that we use in the manuscript.

\section{Spinor Realization of Massless and Massive UIRs of Poincar\'e Algebra}
\label{sec: UIR}

In this work, we consider two types of UIRs of the Poincar\'e algebra.
The first one is the massless helicity representation
($P^2=0$) with  little group $SO(d-2)$,
and  the other one is the massive representation
($P^2<0$ in the mostly positive signature metric)
with little group $SO(d-1)$\,.
These are the only UIRs with 
a finite number of degrees of freedom having positive-definite energies.

Typically, these representations are realized in a 
way that the Poincar\'e covariance is not manifest because
 a certain reference momentum $P_\m=p_\m$ must be used to fix the little group.
However in $d=4$ dimensions, we can realize these representations
in a manifestly covariant fashion by
making use of the spinor representation of the Lorentz algebra.
The price to pay for the covariance is that the representation 
should be realized on a projective space.
In the following, we shall review this realization,
first for the massless UIRs, then for the massive ones.

Before moving to such details,
let us first fix the basic conventions used in this paper.
To construct the Weyl representation of the Lorentz algebra, we use
the Pauli matrices $\sigma^i$ to build the following combinations,
\begin{gather}
	\left(\s^\m\right)_{a\dot b}=\left(1,\vec{\s}\right)_{a\dot b}\,,\qquad
	\left(\bar\s^\m\right)^{\dot ab}=
	\e^{\dot a\dot d}\,\e^{bc}\,
		\left(\s^\m\right)_{c\dot d}=\left(1,-\vec{\s}\right)^{\dot ab}\,,\\
	\left(\s^{\m\n}\right)_a{}^b=
	\frac14\left(\s^\m\,\bar{\s}^\n-\s^\n\,\bar{\s}^\m\right)_a{}^b\,,\qquad
	\left(\bar\s^{\m\n}\right)^{\dot a}{}_{\dot b}=-\frac14\left(\bar\s^\m\, \s^\n-\bar\s^\n\, \s^\m\right)^{\dot a}{}_{\dot b}\,,
\end{gather}
Here, the spinor indices are lowered and raised as
\be
	\psi_a=\e_{ab}\,\psi^{b}\,\,,
	\qquad \psi^a=\e^{ab}\,\psi_{b}\,,
	\qquad\quad \e^{ac}\,\e_{cb}=\d^a_b\,,
\ee
and equivalently for dotted indices.
The matrices $\s^\m$, $\s^{\m\n}$ and $\bar\s^{\m\n}$ can be used to express the components
of any vector $v_\mu$ and anti-symmetric tensor $w_{\m\n}$ in terms of spinorial ones as
\be
	v^\mu=-\frac12\left(\bar\s^\m\right)^{\dot b a}\,v_{a\dot b}\,,
	\qquad
	w^{\m\n}
	=\frac12\,w_{ab}\left(\s^{\m\n}\right)^{ab}
	+\frac12\,\bar w_{\dot a\dot b}\left(\bar\s^{\m\n}\right)^{\dot a\dot b}\,,
\ee
and vice-versa:
\be
	v_{a\dot b}=\left(\s^\m\right)_{a\dot b}\,v_\m\,,\qquad
	w_{ab}=\left(\s^{\m\n}\right)_{ab}\,w_{\m\n}\,,\quad
	\bar{w}_{\dot a \dot b}=\left(\bar\s^{\m\n}\right)_{\dot a\dot b}w_{\m\n}\,.
\ee
When $v_\m$ and $w_{\m\n}$ are real,
the spinorial components should satisfy
${v_{a\dot b}}^*=v_{b\dot a}$ and ${w_{ab}}^*=\bar w_{\dot b\dot a}$\,.
Henceforth, we shall use only these spinorial components
for the Poincar\'e generators.

\subsection{Massless Representations}

Let us begin the discussion with the massless representation.
By making use of a Weyl spinor $\l_{a}$\,,
one can realize the following representation of the Poincar\'e algebra $\iso(3,1)$\,:
\be
	P_{a\dot b}=\lambda_{a}\,\tilde \lambda_{\dot b}\,,\qquad
	L_{ab}=
	\lambda_{(a}\,\frac{\partial}{\partial \lambda^{b)}}\,,\qquad
	\tilde L_{\dot a\dot b}=
	\tilde\lambda_{(\dot a}\,\frac{\partial}{\partial \tilde\lambda^{\dot b)}}\,,
	\label{spinor helicity}
\ee
where we should understand that $\tilde{\lambda}$
and $\lambda$ are complex-conjugate of each other, as it corresponds to real momenta.
Later on, we shall analytically extend them to complex values.
From the vanishing of the quadratic Casimir:
\be
	P^{2}=0\,,
\ee
one can see that \eqref{spinor helicity} is a massless representation.
One can notice that the $\spl_2$'s
generated by $L_{ab}$ and $\tilde L_{\dot a \dot b}$
are in fact in the Schwinger representation, hence
each of them commutes respectively with their \textit{number operator},
\be
	N=\lambda_{a}\,\frac{\partial}{\partial \lambda_{a}}\,,
	\qquad 
	\tilde N=\tilde \lambda_{\dot a}\,\frac{\partial}{\partial\tilde \lambda_{\dot a}}\,.
\ee
Given the form of the translation generator $P_{a\dot b}$\,, only 
the linear combination
\be
	H=N-\tilde N\,,
	\label{H}
\ee
commutes with all the generators of the Poincar\'e algebra.
Therefore, the representation space $V=Fun(\mathbb C^2)$ can be  block-diagonalized with respect to $H$ as
\be
	V=\bigoplus_{h\in \mathbb Z}\,V_{h}\,,
\ee
where the space $V_h$, isomorphic to $\mathbb{C}^2/U(1)$, is given by
\be
\label{Vh}
	V_{h}=\left\{\ 
	f(\lambda, \tilde \lambda)\ \Big|\ 
	\forall\, e^{i\theta}\in U(1),\ 
	f(e^{i\theta}\,\lambda, e^{-i\theta}\,\tilde\lambda)
	=e^{-ih\theta}\,f(\lambda,\tilde \lambda)
	\ \right\}\,. 
\ee
Each space $V_h$ should still carry a faithful representation of Poincar\'e algebra
as $H$ commutes with $\mathfrak{iso}(3,1)$.
In fact,  $V_h$ carries the massless helicity $h$ representation
because $H$ coincides with the helicity operator:
\be
	\frac{W_0}{P_{0}}
	=
	\frac{(\sigma_{0})^{a\dot b}\,
	W_{a\dot b}}
	{(\sigma_{0})^{a\dot b}\,P_{a\dot b}}
	=\frac 12\,H\,,
\ee
where $W_{a\dot b}$ is the Pauli-Lubanski vector,
\be
	W_{a\dot b}
	=P^{c}{}_{\dot b}\,L_{ac}-P_{a}{}^{\dot c}\,\tilde L_{\dot b\dot c}\,.
	\label{PL}
\ee
After analytic continuation,
the $U(1)$ coset condition will be replaced by
the one,
$f(\Omega\,\lambda, \Omega^{-1}\,\tilde\lambda)
	=\Omega^{-h}\,f(\lambda,\tilde \lambda)$
for an arbitrary element $\O$ in $\mathbb C\backslash\{0\}$\,.

\subsection{Massive Representations}
\label{massreps}

Now let us consider the tensor-product of $n$ copies of the representation~\eqref{spinor helicity}:
\be
	P_{a\dot b}=\lambda^I{}_{a}\,\tilde \lambda_{I\,\dot b}\,,\qquad
	L_{ab}=
	\lambda^I{}_{(a}\,\frac{\partial}{\partial \lambda^{I\,b)}}\,,\qquad
	\tilde L_{\dot a\dot b}=
	\tilde\lambda_{I(\dot a}\,\frac{\partial}{\partial \tilde\lambda_I{}^{\dot b)}}\,.
	\label{massive Poincare}
\ee
where $I=1,\ldots, n$\,.
Since each copy is a Poincar\'e UIR,
their tensor products also carry unitary representations under Poincar\'e group,
but reducible ones.
We can reduce this representation into smaller ones by imposing 
certain conditions compatible with the Poincar\'e action.
In this way, we may end up with an irreducible representation.
Appropriate conditions  can be found using differential operators acting on the spinor space
which commute with those in~\eqref{massive Poincare}, which realize the Poincar\'e generators.
We first note that the Lorentz generators commute with two copies of $\gl_n$\,:
\be
	N^I{}_J=\lambda^I{}_{a}\,\frac{\partial}{\partial \lambda^J{}_{a}}\,,
	\qquad
	\tilde N_J{}^I=
	\tilde\lambda_{J\dot a}\,\frac{\partial}{\partial\tilde 
	\lambda_{I\,\dot a}}\,,
\ee
which are in fact  the centralizers of $\spl_2$'s (generated by $L_{ab}$ and $\tilde L_{\dot a\dot b}$) within $\gl_{2n}$'s (generated by $\l^{I}_{a}\frac{\partial}{\partial\l^{J}_{b}}$ and $\tilde\l^{I}_{\dot a}\frac{\partial}{\partial\tilde\l^{J}_{\dot b}}$).
If we extend $\gl_{2n}$ --- that is, any differential operators bilinear in $\l$ and $\partial/\partial \l$
--- to $\syp_{4n}$ by including the operators of the type $\l\,\l$ and $\frac{\partial^2}{\partial \l\,\partial\l}$'s, then
the centralizers are extended with the antisymmetric $n\times n$ tensors
\be
	M^{IJ}=\l^{I}_{\ a}\,\l^{Ja}\,, \qquad \tilde M_{IJ}=	\tilde\l_{I\dot a}\,\tilde\l_{J}{}^{\dot a}\,,
\ee
and their $\l\leftrightarrow\partial/\partial\l$ conjugates. The latter
commute with $M^{IJ}$ and $\tilde M_{IJ}$, however they have non-trivial commutators with the translation generators.
Moreover among $N^I{}_J$ and $\tilde N_J{}^I$\, only the combination
\be
	K^{I}_{J}=N^I{}_J-\tilde N_J{}^I\,,\qquad 
\ee
commutes with the whole Poincar\'e algebra.
Therefore,
we can decompose the spinor space $\mathbb{C}^{2n}$ in terms of the UIRs of 
the algebra, 
to which we shall refer to as $\mathfrak{A}_n$\,, generated by $K^{I}_{J}$, $M^{IJ}$ and $\tilde M_{IJ}$\,:
it is a semi-direct sum of $\uu(n)$ (generated by $K^{I}_{J}$) and the antisymmetric tensor product 
of  fundamental (for $M^{IJ}$) and anti-fundamental (for $\tilde M_{IJ}$) representations.
By choosing a particular UIR of $\mathfrak{A}_n$,
we can reduce the tensor product 
representation of Poincar\'e algebra into a smaller one.
As in the Poincar\'e case, the UIRs of $\mathfrak{A}_n$ can be 
classified according to the  value of the quadratic Casimir,
\be
\label{C2}
	 C_2(\mathfrak{A}_n)=-\frac12\,M^{IJ}\,\tilde M_{IJ}\,,
\ee
which in fact coincides with that of the Poincar\'e algebra:
\be
	C_2(\iso(3,1))=P^2=-\frac12\,P_{a\dot b}\,P^{a\dot b}
	= C_2(\mathfrak{A}_n)\,.
	\label{mass shell}
\ee
 Depending on this value,
 we may classify the representations of $\mathfrak{A}_n$.

\subsubsection*{Case $n=2$}

For the massive representations, 
it will be sufficient to consider the $n=2$ case,
where we have only one component for each of $M^{IJ}=-\e^{IJ}\,M$ and $\tilde M_{IJ}=\e_{IJ}\,\tilde M$\,. The latter generators commute with the $\su(2)\subset \uu(2)$ generated by $\cK^I_J$\,:
\ba
\label{cKs}
	&\cK^I_J=K^I_J-\frac12\,\delta^I_J\,K_K^K
	=\cN^I{}_J-\tilde\cN_J{}^I\,,\\
	&
	\cN^I{}_J=N^I{}_J-\frac12\,\delta^I_J\,N^K{}_K\,,
	\qquad
	\tilde \cN_I{}^J=N_I{}^J-\frac12\,\delta^I_J\,\tilde N_K{}^K\,,
\ea
whereas the $\uu(1)$ {part $K=K^I{}_I$} satisfies the commutation relations:
\be
	[K,M]=2\,M\,,\qquad [K,\tilde M]=-2\,\tilde M\,.
\ee
We see that $K,M,\tilde M$ form an $\iso(2)$ algebra. Hence, we see that $\mathfrak{A}_2\simeq 
\su(2)\oplus \iso(2)$\,.
It turns out that the determination of an irreducible representation under $\iso(2)$ 
fixes the mass value of the Poincar\'e representation,
while choosing an irreducible representation for $\su(2)$ fixes
the spin. We can therefore associate 
the massive little group $SO(3)$ of the Poincar\'e group
with the $\su(2)$ generated by $\cK^I_J$\,.
One of the simplest ways to see this is to compute 
the corresponding Casimir operators. The quadratic Casimir $P^2$ of Poincar\'e algebra is already
fixed in this case by~\eqref{mass shell}, and given by
$C_2(\mathfrak{A}_2)=C_2(\iso(2))=-M\,\tilde M$\,.
Similarly to $P^2$\,, the square of Pauli-Lubanski vector~\eqref{PL} commutes with all the other generators.
After a simple manipulation, we get
\be 
	W^2=-\frac12\,W_{a\dot b}\,W^{a\dot b}
	=-\frac14\,P_{a\dot b}\,P^{a\dot b}\left(L_{cd}\,L^{cd}+\tilde L_{\dot c\dot d}\,\tilde L^{\dot c\dot d}\right)
	+P^{a\dot c}\,P^{b\dot d}\,L_{ab}\, \tilde L_{\dot c\dot d}\,,
\ee
which can be further simplified using the identities, valid only for $n=2$,
\ba
	& L_{ab}\,L^{ab}=\cN^I{}_J\,\cN^J{}_I\,,
	\qquad
	\tilde L_{\dot a\dot b}\,\tilde L^{\dot a\dot b}=
	\tilde \cN_I{}^J\,\tilde \cN_J{}^I\,,\\
&
	P^{a\dot c}\,P^{b\dot d}\,L_{ab}\, \tilde L_{\dot c\dot d}
	=M\,\tilde M\,\cN^I{}_J\,\tilde \cN_I{}^J\,.
\ea
Finally, combining  the above formulas, we can show that
\be
	W^2=-\frac12\,M\,\tilde M\,\cK^I_J\,\cK^J_I\,.
\ee
This makes it clear that when $P^2=-M\,\tilde M=-m^2<0$\,, by taking the `spin $s$' representation of $\su(2)$\,,
the corresponding Poincar\'e representation becomes also that of massive spin~$s$\,.
To recapitulate, starting from $V=Fun(\mathbb C^4)$\,,
we get
\be
	V=\bigoplus_{m\in \mathbb R,\, s \in \mathbb N/2}\,V_{m,s}\,,
\ee
where $V_{m,s}$, isomorphic to $\mathbb{C}^4/\mathfrak{A}_2$, is given by
\be
\label{Vms}
	V_{m,s}=\left\{\ 
	f_{r,h}(\l,\tilde \l)\ \Bigg|\ 
	\begin{array}{c}
	\forall\, g \in SU(2)\,,\ 
	f_{r,h}(g\,\lambda, g^{-1}\,\tilde\lambda)
	=D^s_{hh'}(g)\,f_{r,h'}(\lambda,\tilde \lambda)\\
	\forall\, e^{i\,\theta} \in U(1)\,,\ f_{r,h}(e^{i\,\theta}\,\l,e^{-i\,\theta}\tilde\l)=
	e^{2\,i\,r\,\theta}\,f_{r,h}(\l,\tilde \l)\\
	\l_{I\,a}\,\l^{I\,a}\,f_{r,h}(\l,\tilde \l)=2\,m\,f_{r+1,h}(\l,\tilde \l)\\
	\tilde \l_{I\,\dot a}\,\tilde\l^{I\,\dot a}\,f_{r,h}(\l,\tilde \l)=2\,m\,f_{r-1,h}(\l,\tilde \l)\
	\end{array}	
	\ \right\},
\ee
where $D^s_{hh'}(g)$ is the Wigner D-matrix
and $\lambda_I=\e_{IJ}\,\l^J$ and $\tilde \l^I=\e^{IJ}\,\tilde\l_J$. Since this matrix does not depend on the label $r$, representations with different
values of $r$ can be considered physically equivalent. We will make use of this fact later on in Section~\ref{sec: massive}.
See Appendix \ref{app}
for detailed account
of how~\eqref{Vms} is related to
the standard massive representation.

\section{Massless Interactions}
\label{sec: massless}

We will match now each local cubic interaction to one of the three-point amplitudes
classified using the spinor-helicity method.
We begin this analysis with the case
where all three fields or representations are massless.
We restrict to the case of bosonic fields for the sake of simplicity.
Before making an explicit link between them,
we review the classification of 
massless three-point amplitudes in spinor-helicity variables
and the construction of gauge-invariant local cubic-interaction vertices and derive parity odd cubic vertices in covariant form.

\subsection{Three-point Amplitudes}

Let us denote the asymptotic states of the three massless particles by
\be
	\lvert\lambda^I,\tilde\lambda_I;h_I\rangle\,, \quad I=1,2,3\,.
\ee
As we discussed in Section~\ref{sec: UIR}, each asymptotic state furnishes a representation of $\iso(3,1)$.
The three-particle amplitude carries information about this representation in the sense
of being a function living in the spaces $V_{h_I}$ defined in~\eqref{Vh}. A way to re-state this is by saying
that the helicity operator in~\eqref{H} acts on the amplitude as it acts on the one-particle states. This fact essentially
determines the three-point amplitude up to a coupling constant \cite{Benincasa:2007xk}. Discarding
delta-function contributions (apart from the momentum-conserving delta-function, that we omit
in what follows), the solution to the differential equations that follow from applying
the three helicity operators to the amplitude is
\be
\label{M3f}
	M_3^{h_1,h_2,h_3}=\kp12^{h_3-h_1-h_2}\kp31^{h_2-h_3-h_1}\kp23^{h_1-h_2-h_3}
	f\left(\kp{I}{J}\bp{I}{J}\right)\,,
\ee
with  $\la I,J\ra=\l^{I}_{\ a}\,\l^{Ja}$ and  $[I,J]=
\tilde\l_{I\,\dot a}\,\tilde \l_{J}^{\dot a}$.
Here, $f$ is an unknown function that can be determined with three physical requirements.
  One is that
the amplitude should not be singular. Another is momentum conservation, that implies that either
$\kp{I}{J}=0$ or $\bp{I}{J}=0$. The remaining one is the fact that whenever $h_1+h_2+h_3\neq0$, the
amplitude must vanish on the real sheet\footnote{%
In the case $h_1+h_2+h_3=0$, the amplitude does not need to vanish a priori for real momenta,
which can line up along a null direction. However, no consistent interactions are known of this type.
We will comment on this point again at the end of this Section~\ref{sec: massless}.
 }. %
This gives
\be
\label{M3g}	M_3^{h_1,h_2,h_3}=\begin{cases}
	g_\textrm{\tiny H}\,\kp12^{h_3-h_1-h_2}\,\kp31^{h_2-h_3-h_1}\,\kp23^{h_1-h_2-h_3} & \textrm{when }h_1+h_2+h_3<0\\
	g_\textrm{\tiny A}\,\bp12^{h_1+h_2-h_3}\,\bp31^{h_3+h_1-h_2}\,\bp23^{h_2+h_3-h_1} & \textrm{when }h_1+h_2+h_3>0
	\end{cases}\,.
\ee
The coupling constants $g_\textrm{\tiny H}$, $g_\textrm{\tiny A}$ above\footnote{%
The subindices {\scriptsize H} and {\scriptsize A} stand for holomorphic and anti-holomorphic, referring
to their dependence on the $\lambda$ and $\tilde\lambda$ spinors respectively.}
are unrelated in a theory with no well-defined parity. For parity-even or -odd theories, they are equal up to a sign . 
 Let us remark
here that if we drop the non-singular requirement, we could formally have singular amplitudes
obeying~\eqref{M3f} (recall that $\kp{I}{J}\bp{I}{J}=0$ by momentum conservation). This can
make sense in certain contexts \cite{Bern:2005hs}, and we will see at the end of this section another example
of this happening.

Finally note that there is no restriction on the spin of the scattering particles, which can be
arbitrarily large. For given spins $s_1$, $s_2$ and $s_3$, there are generically four types
of amplitudes associated: 
\be
\label{h1h2h3}
	(h_1,h_2,h_3)\in\left\{\left(\pm s_1, \pm s_2, \pm s_3\right),\left(\mp s_1, \pm s_2, \pm s_3\right),
	\left(\pm s_1, \mp s_2, \pm s_3\right),\left(\pm s_1, \pm s_2, \mp s_3\right)\right\},
\ee
which we grouped in parity-conjugated pairs.

\subsection{Cubic vertices}
\label{3ptV}

Let us now turn to the Lagrangian description of cubic interactions. Local and gauge-invariant
cubic Lagrangians have been completely classified in any $D$-dimensional space-time ($D\ge4$).
We briefly review here the derivation of covariant cubic vertices in four dimensions. 

As is common practice, to deal with arbitrary higher-spin fields we introduce auxiliary variables $u^\mu$
and define the generating functions:
\be
\label{vphi}
	\varphi(x,u)=
	\sum_{s=0}^{\infty}\frac{1}{s!}\,\varphi_{\mu_{1}\cdots\mu_{s}}(x)\,
	u^{\mu_{1}}\cdots u^{\mu_{s}}\,.
\ee
The massless system requires gauge symmetries 
in order to propagate  the correct number of on-shell degrees of freedom,
and the gauge transformation takes the following form:
\be
\label{gauge0}
	\delta\varphi(x,u)=u\cdot\partial_{x}\,\varepsilon(x,u)+\cdots\,,
\ee
where the dots contain terms of higher order in the number of fields (namely they contain the non-linear
part of the gauge transformation), which are not needed at cubic level. Indeed, 
from the gauge invariance of the full action, it can be shown that the generic cubic vertex
\be
\label{gen cub}
	S^{\sst (3)}=\int d^{4}x\,C	(\partial_{x_{I}},\partial_{u_{I}})\,
	\varphi^{1}(x_{1},u_{1})\,\varphi^{2}(x_{2},u_{2})\,
	\varphi^{3}(x_{3},u_{3})\,
	\Big\rvert_{\substack{x_{I}=x \\ u_{I}=0}}\,,
\ee
should satisfy
\be
\label{dC}
	\left[C(\partial_{x_{J}},\partial_{u_{J}}),u_I\cdot\partial_{x_I}\right]\approx0\,,\quad I=1,2,3\,,
\ee
where $\approx$ means modulo the Frondsal equations of motion \cite{Fronsdal:1978rb}.
In order to solve equation~\eqref{dC}, one has to analyze what are the variables that $C$ can
possibly depend on. Since we will only require the on-shell content of the vertex later on, we
can just focus on its transverse and traceless (TT) part, that we denote by $C^\textrm{\tiny TT}$,
and which also satisfies equation~\eqref{dC}
where $\approx$ means now modulo the Fierz system:
\be
	\partial_{x_I}^2\,\varphi^I\approx0\,,
	\qquad
	\partial_{x_I}\cdot\partial_{u_I}\,\varphi^I\approx0\,,
	\qquad
	\partial^2_{u_I}\varphi^I\approx 0\,.
	\label{Fierz}
\ee
In order to continue the analysis, we need to distinguish 
the cases where the vertices involve a
Levi-Civita epsilon tensor (hence parity-odd)
or not (parity-even).

\subsubsection{Parity-even vertices}

Let us begin with the parity-even cases.
Since the vertices do not involve any $\e_{\m\n\r\s}$ tensor,
the vertex function $C^\textrm{\tiny TT}$ can only
depend on the six variables,
\be
\label{y and z}
	Y_{I}=\partial_{u_{I}}\!\cdot\partial_{x_{I+1}}\,,\qquad
	Z_{I}=\partial_{u_{I+1}}\!\!\cdot\partial_{u_{I-1}}\,.
\ee
$C^\textrm{\tiny TT}$ is then easily determined using the commutators,
\be
\label{commus}
	\left[Y_I,u_J\cdot\partial_{x_J}\right]=0\,,\quad
	\left[Z_I,u_I\cdot\partial_{x_I}\right]=0\,,\quad
	\left[Z_I,u_{I\pm1}\cdot\partial_{x_{I\pm1}}\right]=\mp Y_{I\mp1}\,.
\ee
The solution to the equations~\eqref{dC} is
\be
\label{3flat}
	C^\textrm{\tiny TT}=\sum_{n=0}^{s_1}\,\lambda_{n}^{(s_1,s_2,s_3)}\,G^n\,Y_1^{s_1-n}\,Y_2^{s_2-n}\,Y_3^{s_3-n}\,,
\ee
where the \mt{\lambda_{n}^{(s_1,s_2,s_3)}}'s are independent coupling constants that
ought to be fixed by the quest for consistency of higher order interactions, and we defined
$G$ as the combination,
\be
	G=Y_{1}\,Z_{1}+Y_{2}\,Z_{2}+Y_{3}\,Z_{3}\,.
\ee
The discussion up to here is actually valid in any space-time dimension. In
four dimensions the variables $Y_I$ and $Z_I$ are not independent, as generalized Kronecker-delta
identities in four dimensions\footnote{
More explicitly, we should consider the identity
$\delta^{\nu_1\nu_2\nu_3\nu_4\nu_5}_{\mu_1\mu_2\mu_3\mu_4\mu_5}
\partial_{u_1}^{\mu_1}\partial_{u_2}^{\mu_2}\partial_{u_3}^{\mu_3}\partial_{x_1}^{\mu_4}\partial_{x_2}^{\mu_5}
\partial_{u_1^{\nu_1}}\partial_{u_2^{\nu_2}}\partial_{u_3^{\nu_3}}\partial_{x_1^{\nu_4}}\partial_{x_2^{\nu_5}}=0$.
 Hereupon we will refer to these sort of identities as Schouten identities.
} imply $Y_1\,Y_2\,Y_3\,G\approx0$.
This makes the expression~\eqref{3flat} collapse to just two possible parity-even
vertices:
\be
\label{CPP}
	C^\textrm{\tiny TT}= g_\textrm{\tiny min}\,G^{s_{1}}\,Y_{2}^{s_{2}-s_{1}}\,Y_{3}^{s_{3}-s_{1}}
	+g_\textrm{\tiny non}\,Y_{1}^{s_{1}}\,Y_{2}^{s_{2}}\,Y_{3}^{s_{3}}\,,
\ee
where we assumed $s_1\le s_2 \le s_3$ without loss of generality.
The first vertex is the minimal coupling and contains $s_3+s_2-s_1$ derivatives. The second
one contains $s_1+s_2+s_3$ derivatives instead, and it is usually called non-minimal
coupling. Notice that these two vertices coincide when $s_1=0$.

\subsubsection{Parity-odd vertices}

The analysis of parity-odd cubic vertices is analogous to the parity-even case, except that now
\be
	C^\textrm{\tiny TT}
	=\sum_{I=1}^3\,V_I\,F^{\sst (V)}_I(Y,Z)+W_I\,
	F^{\sst (W)}_I(Y,Z)\,,
	\label{C V W}
\ee
depends linearly on the variables
\be
\label{VW}
	V_I=\epsilon^{\m\n\r\s}\partial_{u_{I+1}^{\m}}\partial_{x_{I+1}^\n}\partial_{u_{I-1}^{\r}}\partial_{x_{I-1}^\s}\,, \qquad
	W_I=\epsilon^{\m\n\r\s}\partial_{u_1^{\m}}\partial_{u_2^{\n}}\partial_{u_3^{\r}}\partial_{x_I^\s}\,.
\ee
It is important to note here that the expression~\eqref{C V W}
contains in general a redundancy because
the six variables $V_I$'s and $W_I$'s are not independent --- using Schouten identities and momentum conservation we get the following six relations:
\be
\label{WVcons}
\begin{aligned}
	W_I\,Y_I&\approx V_{I+1}Z_{I-1}+V_{I-1}Z_{I+1}\,,\quad	&	W_1+W_2+W_3&\approx0\,,\\
	V_1\,Y_1&\approx V_2\,Y_2\approx V_3\,Y_3\,,		&	V_I\,Y_I\,G&\approx0\,.
\end{aligned}
\ee
The redundancy can be removed by expressing $V_2, V_3$ and $W_I$'s
in terms of the other variables as
\be
\label{V W rep}
	V_I\approx V_1\frac{Y_1}{Y_I}\,,\;\;
	W_I\approx V_1\frac{Y_{I+1}Z_{I+1}+Y_{I-1}Z_{I-1}}{Y_2\,Y_3}\,,\;\;
	I=2,3\quad;\quad
	W_1\approx-V_1\,\frac{G+Y_1Z_1}{Y_2\,Y_3}\,.
\ee
Since these relations commute with the gauge variations,
removing the redundancy before solving the gauge invariance condition~\eqref{dC}
yields a simpler expression for the vertex, namely $C^\textrm{\tiny TT}=V_1\,F(Y,Z)$. 
Notice however that since the replacements~\eqref{V W rep}
involve negative powers of $Y_2$ and $Y_3$, 
the function $F$ is allowed to have terms proportional to the negative
powers $Y_2^{-1}$, $Y_3^{-1}$ or $(Y_2Y_3)^{-1}$. 
Negative powers of $Y_I$ do not make sense as it would mean a 
negative number of contractions, but
actually they might be just an artifact of our procedure, which is purposed to remove redundancies. 
As we will show below, it is possible that even when these negative powers show up, the vertex
still admits a polynomial expression if the relations~\eqref{V W rep} can be inverted.

Given that $[V_1,u_I\cdot\partial_{x_I}]=0$, we can immediately see that there are only two possible
gauge-invariant parity-odd vertices:
\be
\label{CPV}
	C^{\rm\sst TT}= g_\textrm{\tiny min,PO}\,V_1\,G^{s_1}\,Y_{2}^{s_{2}-s_1-1}\,Y_{3}^{s_3-s_1-1}
	+g_\textrm{\tiny non,PO}\,V_1\,Y_{1}^{s_1}\,Y_{2}^{s_2-1}\,Y_{3}^{s_3-1}\,.
\ee
They respectively have $s_2+s_3-s_1$ and $s_1+s_2+s_3$ derivatives,
hence can be naturally paired with the parity-even minimal and non-minimal coupling vertices.
For this reason, we also refer to these parity-odd vertices as
minimal and non-minimal couplings. Let us note that, similarly to parity-even case, the two vertices coincide for $s_1=0$.
Notice  that the minimal-coupling vertex has negative powers of $Y_I$'s
when \mt{s_1=s_2}\,.
However, for $s_2 < s_3$\,,
it can be brought to a polynomial form
using the identities~\eqref{WVcons} as
\be
\label{PVspecial}
	V_1\,G^{s_1}\,Y_{2}^{-1}\,Y_{3}^{s_3-s_1-1}\approx\frac12\left[V_1\,Z_2-V_2\,Z_1+
	\left(W_2-W_1\right)Y_3\right]G^{s_1-1}\,Y_{3}^{s_3-s_1-1}\,,
\ee
where we have chosen the  form symmetric under exchange of 1 and 2 among various equivalent expressions.
About the case with $s_1=s_2=s_3$,  
it is impossible to remove completely the negative powers of
$Y_2$ and $Y_3$ using~\eqref{WVcons} from the minimal-coupling vertices.
Therefore we conclude that, compared to the parity-even vertices,
the parity-odd vertices miss the minimal ones with all equal spins.
This result is compatible with the work \cite{Boulanger:2005br}
on spin-three parity-odd interactions, where the vertex was found only for 
three and five dimensions.  

There is another case of coincident spins: $s_1<s_2=s_3$, for which the equation~\eqref{CPV} does not contain negative powers of $Y_I$'s. In this case, the vertex operator has symmetry property with respect to exchange of second and third fields, up to a factor $(-1)^{s_1}$, which suggests, similarly to parity-even cases, to include Chan-Paton structures in the case of odd $s_1$. Instead in the case of $s_1=s_2<s_3$, the vertex~\eqref{PVspecial} has the opposite property --- Chan-Paton structures are needed for even $s_3$. This strange difference suggests intuitive understanding of the case $s_1=s_2=s_3$, which, belonging to both of the above classes of vertices, should have both symmetric and antisymmetric properties with respect to exchange of any two fields, which cannot be satisfied by any non-vanishing vertex operator.

\subsection{Match}
\label{match}

We will investigate here if the amplitudes~\eqref{M3g} match the ones obtained from the cubic
vertices~\eqref{CPP} and~\eqref{CPV}. For that it suffices to translate the Lagrangian vertices into
amplitudes.

Given the form of the vertex~\eqref{gen cub}, Feynman rules instruct us to extract the coefficient of
$\prod_{I=1}^3\prod_{k=1}^{s_I}\partial_{u_I^{\mu_k}}$ of $C(\partial_{x_I},\partial_{u_I})$, multiply
it by $(-i)\,e^{i(p_1+p_2+p_3)\cdot x}$, and contract it with the polarization tensors of the three external
particles. This is equivalent to writing
\be
\label{M3C}
	\widetilde{M}_3=\int d^{4}x\,C^{\textrm{\tiny TT}}(G,Y_i,V_1)\,
	\varphi_\textrm{\tiny O-S}^{1}(x_{1},u_{1})\,\varphi_\textrm{\tiny O-S}^{2}(x_{2},u_{2})\,
	\varphi_\textrm{\tiny O-S}^{3}(x_{3},u_{3})\,
	\Big\rvert_{\substack{x_{I}=x \\ u_{I}=0}}\,,
\ee
where $\widetilde{M}_3$ is the formal sum of the position-space amplitudes for all helicity
configurations and all spins, and the {\scriptsize O-S} subindex refers to the fact that the $\varphi^I$ should
satisfy the equations of motion. The equivalence holds because the on-shell evaluation of 
the massless higher-spin fields yields polarization tensors times plane-wave exponentials:
\be
\label{phiOS}
	\varphi_\textrm{\tiny O-S}(x,u)=\sum_{s=0}^\infty\frac{1}{s!}\,
	\int d^4p\,\left(\varphi^-(p)\,\epsilon^-_{\mu_{1}\cdots\mu_{s}}(p)+\varphi^+(p)\,\epsilon^+_{\mu_{1}\cdots\mu_{s}}(p)\right)\,
	u^{\mu_{1}}\cdots u^{\mu_{s}}\,e^{ip\cdot x}\,.
\ee
Therefore, let us start by solving the on-shell conditions~\eqref{Fierz} for a generic
$\varphi(x,u)$. We will do so in terms of spinor variables to connect with~\eqref{M3g}.
For that purpose, it is quite convenient to use light-cone variables:
\be
\begin{aligned}
	x^{\pm}&=x^{0}\pm x^{3}\,,	&	 z&=x^{1}+i\,x^{2}\,,\\
	u^{\pm}&=u^{0}\pm u^{3}\,,	&	 \omega&=u^{1}+i\,u^{2}\,.
\end{aligned}
\ee
We start by fixing the gauge. We impose the light-cone gauge condition,
\be
\label{LC}
	\partial_{u^{-}}\,\varphi(x^{+},x^{-},z,\bar z\,;\,u^{+},u^{-},\omega,\bar\omega)=0\,,
\ee
which simply implies that $\varphi$ cannot depend on $u^{-}$. In this gauge, the transverse condition reads
\be
	(-\partial_{u^{+}}\partial_{x^{-}}+\partial_{z}\,\partial_{\bar\omega}
	+\partial_{\bar z}\,\partial_{\omega})\,
	\varphi(x^{+},x^{-},z,\bar z\,;\,u^{\ed{+}},\omega,\bar\omega)=0\,,
\ee
which can be easily solved by
\be
\label{lc fix}
	\varphi(x^{+},x^{-},z,\bar z\,;\,u^{\ed{+}},\omega,\bar\omega)=
	\exp\left[\frac{u^{+}}{\partial_{x^{-}}}\,(\partial_{z}\,\partial_{\bar\omega}
	+\partial_{\bar z}\,\partial_{\omega})\right]
	\varphi_{\rm\sst l.c.}(x^{+},x^{-},z,\bar z\,;\,\omega,\bar\omega)\,,
\ee
where the subindex {\small l.c.} refers to light-cone.
The next condition to solve is the d'Alembertian equation, \mt{\Box\,\varphi_{\rm\sst l.c.}=0}, whose solution is a simple
superposition of plane waves: 
\begin{equation}
\label{spinor}
	\varphi_{\rm\sst l.c.}(x,\o,\bar\o)\!=\!
	\int\! \frac{d^{2}\lambda\, d^{2} \tilde \lambda}{{\rm vol}(GL(1))}\,
	e^{i\left(\frac i2\,x^{a\dot a}\,\lambda_{a}\,\tilde \lambda_{\dot a}\right)}
	\exp\left[\frac{\tilde\lambda_{\dot2}}{\lambda_{2}}\, \omega\,\partial_{\chi}+
	\frac{\lambda_{2}}{\tilde\lambda_{\dot2}}\, \bar\omega\,\partial_{\bar \chi}
	\right]
	\phi(\lambda,\tilde \lambda\,;\,
	\chi,\bar \chi)\,\Big|_{\chi=\bar \chi=0}\,,
\end{equation}
where we have decided to introduce the second exponential so that the ``wave-function''
$\phi$ carries helicity representations: 
\be
	\phi(\Omega\,\lambda,\Omega^{-1}\,\tilde \lambda\,;\,
	\Omega^{-2}\,\chi,\Omega^{2}\,\bar \chi)
	=\phi(\lambda,\tilde \lambda\,;\,
	\chi,\bar \chi)
	\qquad 
	[\Omega \in \mathbb C]\,.	
\ee
Note that the integrand above becomes singular when the second components of the spinors are vanishing.
This just occurs because we assumed that $\partial_{x^-}\varphi\neq0$ when imposing the light-cone
condition~\eqref{LC}. This is equivalent to assuming that $\lambda_2\tilde\lambda_{\dot2}\neq0$ .

Now, using the relations~\eqref{lc fix} and~\eqref{spinor},  
the effect of derivatives and contractions on the wave-function $\varphi$ are given by
\be
\label{map}
\begin{aligned}
	\big(\,\partial_{x^{+}},\partial_{x^{-}},\partial_{z},\partial_{\bar z}\,\big)&
	\varphi
	&&\leftrightarrow &&
	i\, \big(-\lambda_{1}\,\tilde\lambda_{\dot1},-\lambda_{2}\,\tilde\lambda_{\dot2},
	\lambda_{1}\,\tilde\lambda_{\dot2},\lambda_{2}\,\tilde\lambda_{\dot2}\,\big)
	\phi\,,\\
	\big(\,\partial_{u^{+}},\partial_{u^{-}},\partial_{\omega},\partial_{\bar \omega}\,\big)&
	\varphi
	&&\leftrightarrow&&
	\left[\big(-\tilde\lambda_{\dot1},0,\tilde\lambda_{\dot2},0\,\big)\,\frac{\partial_{\chi}}{\lambda_{2}}\,
	+\big(-\lambda_{1},0,0,\lambda_{2}\,\big)\,\frac{\partial_{\bar\chi}}{\tilde\lambda_{\dot2}}\,\right]
	\phi\,.
\end{aligned}
\ee
From here we can see that the traceless condition  \mt{\partial_{u}^{2}\,\varphi=0} also takes a very simple form:
\be
	\partial_{\chi}\,\partial_{\bar\chi}\,
	\phi(\lambda,\tilde\lambda\,;\,\chi,\bar\chi)=0
	\quad \Rightarrow \quad
	\phi(\lambda,\tilde\lambda\,;\,\chi,\bar\chi)=
	\phi^{+}(\lambda,\tilde\lambda\,;\,\chi)
	+\phi^{-}(\lambda,\tilde\lambda\,;\,\bar\chi)\,,
\ee
forbidding mixed contractions. Notice that this gives rise to the decomposition~\eqref{phiOS}.
In order to evaluate~\eqref{M3C}, we can just plug~\eqref{lc fix} and~\eqref{spinor} in there, then
use~\eqref{map} to compactly write the resulting expression. Indeed, the operators that can appear
in the vertex cast nicely as
\begin{align}
	-4i\,G &\leftrightarrow&&
	\frac{\bp23^{3}}{\bp12\bp31}\,\partial_{\bar\chi_{1}}\,\partial_{\chi_{2}}\,\partial_{\chi_{3}}+
	\frac{\bp31^{3}}{\bp23\bp12}\,\partial_{\chi_{1}}\,\partial_{\bar\chi_{2}}\,\partial_{\chi_{3}}+
	\frac{\bp12^{3}}{\bp31\bp23}\,\partial_{\chi_{1}}\,\partial_{\chi_{2}}\,\partial_{\bar\chi_{3}}\nn
	\label{G}
	&&&\hspace{-6ex}+\,\frac{\kp23^{3}}{\kp12\kp31}\,\partial_{\chi_{1}}\,\partial_{\bar\chi_{2}}\,\partial_{\bar\chi_{3}}+
	\frac{\kp31^{3}}{\kp23\kp12}\,\partial_{\bar\chi_{1}}\,\partial_{\chi_{2}}\,\partial_{\bar\chi_{3}}+
	\frac{\kp12^{3}}{\kp31\kp23}\,\partial_{\bar\chi_{1}}\,\partial_{\bar\chi_{2}}\,\partial_{\chi_{3}}\,,\\
	\label{Y}
	-2i\,Y_{I} &\leftrightarrow&&
	\frac{\bp{I}{J}\bp{I}{K}}{\bp{J}{K}}\,\partial_{\chi_{I}}+\frac{\kp{I}{J}\kp{I}{K}}{\kp{J}{K}}\,\partial_{\bar\chi_{I}}\,,\\
	\label{V1}
	-4i\,V_1&\leftrightarrow&&
	\bp23^2\partial_{\chi_2}\partial_{\chi_3}-\kp23^2\partial_{\bar{\chi}_2}\partial_{\bar{\chi}_3}\,.
\end{align}
Notice that the expressions above contain singular terms of the form $\frac{0}{0}$ when momentum
conservation is taken into account. The operators $G,Y_I,V_1$ are of course not singular. We have
purposely introduced these singularities by substituting
\be
	\frac{\lambda^J_2}{\lambda^I_2}=-\frac{\bp{I}{K}}{\bp{J}{K}}\,,\qquad
	\frac{\tilde\lambda_{J2}}{\tilde\lambda_{I2}}=-\frac{\kp{I}{K}}{\kp{J}{K}}\,,
\ee
in order to make the final expressions more appealing, and also because this will allow us to make
easier contact with~\eqref{M3f}, where momentum conservation is not explicitly imposed. Using the
formulae~\eqref{G}-\eqref{V1}, we can easily evaluate the cubic vertices. 

Let us start with the non-minimal vertices with
$(s_1+s_2+s_3)$ derivatives. Omitting factors of $2$ and $i$, we have, schematically,
\be
\begin{gathered}
	C^\textrm{\tiny TT}=g_\textrm{\tiny non}\,Y_{1}^{s_{1}}\,Y_{2}^{s_{2}}\,Y_{3}^{s_{3}}+
	g_\textrm{\tiny non,PO}\,V_1\,Y_{1}^{s_{1}}\,Y_{2}^{s_{2}-1}\,Y_{3}^{s_{3}-1}\\
	\updownarrow\\
	\left(g_\textrm{\tiny non}+g_\textrm{\tiny non,PO}\right)
	\bp12^{s_{1}+s_{2}-s_{3}}\,\bp23^{s_{2}+s_{3}-s_{1}}\,\bp31^{s_{3}+s_{1}-s_{2}}\,
	\partial_{\chi_{1}}^{\,s_{1}}\,\partial_{\chi_{2}}^{\,s_{2}}\,\partial_{\chi_{3}}^{\,s_{3}}\\
	+\left(g_\textrm{\tiny non}-g_\textrm{\tiny non,PO}\right)
	\kp12^{s_{1}+s_{2}-s_{3}}\,\kp23^{s_{2}+s_{3}-s_{1}}\,\kp31^{s_{3}+s_{1}-s_{2}}\,
	\partial_{\bar\chi_{1}}^{\,s_{1}}\,\partial_{\bar\chi_{2}}^{\,s_{2}}\,\partial_{\bar\chi_{3}}^{\,s_{\ed{3}}}\,.
\end{gathered}
\ee
In view of equation~\eqref{M3C}, we see that this vertex corresponds to the on-shell amplitude~\eqref{M3g}
in the first helicity configuration of~\eqref{h1h2h3}, identifying the coupling constants as
\be
\label{gids}
	g_\textrm{\tiny H}\sim g_\textrm{\tiny non}-g_\textrm{\tiny non,PO}\,,\qquad
	g_\textrm{\tiny A}\sim g_\textrm{\tiny non}+g_\textrm{\tiny non,PO}\,.
\ee
{When one of the vertices with well-defined parity is not present}, we see that $g_\textrm{\tiny H}$ and $g_\textrm{\tiny A}$
are indeed related. 

For the minimal-coupling vertices with $s_1<s_2\le s_3$, we get
\be
\begin{gathered}
	C^\textrm{\tiny TT}=g_\textrm{\tiny min}\,G^{s_{1}}\,Y_{2}^{s_{2}\ed{-s_1}}\,Y_{3}^{s_{3}\ed{-s_1}}+
	g_\textrm{\tiny min,PO}\,V_1\,G^{s_{1}}\,Y_{2}^{s_{2}\ed{-s_1}-1}\,Y_{3}^{s_{3}\ed{-s_1}-1}\\
	\updownarrow\\
	g_\textrm{\tiny A}
	\frac{\bp23^{s_{1}+s_{2}+s_{3}}}{\bp12^{s_{1}+s_{3}-s_{2}}\bp31^{s_{1}+s_{2}-s_{3}}}\,
	\partial_{\bar\chi_{1}}^{\,s_{1}}\,\partial_{\chi_{2}}^{\,s_{2}}\,\partial_{\chi_{3}}^{\,s_{3}}	+
	g_\textrm{\tiny H}\frac{\kp23^{s_{1}+s_{2}+s_{3}}}{\kp12^{s_{1}+s_{3}-s_{2}}
	\kp31^{s_{1}+s_{2}-s_{3}}}\,
	\partial_{\chi_{1}}^{\,s_{1}}\,\partial_{\bar\chi_{2}}^{\,s_{2}}\,\partial_{\bar\chi_{3}}^{\,s_{3}}\\
	+ g_\textrm{\tiny A}f_2
	\frac{\bp31^{s_{1}+s_{2}+s_{3}}}{\bp12^{s_{2}+s_{3}-s_{1}}\bp23^{s_{1}+s_{2}-s_{3}}}\,
	\partial_{\chi_{1}}^{\,s_{1}}\,\partial_{\bar\chi_{2}}^{\,s_{2}}\,\partial_{\chi_{3}}^{\,s_{3}}	+
	g_\textrm{\tiny H}f_2\frac{\kp31^{s_{1}+s_{2}+s_{3}}}{\kp12^{s_{2}+s_{3}-s_{1}}
	\kp23^{s_{1}+s_{2}-s_{3}}}\,
	\partial_{\bar\chi_{1}}^{\,s_{1}}\,\partial_{\chi_{2}}^{\,s_{2}}\,\partial_{\bar\chi_{3}}^{\,s_{3}}\\
	+g_\textrm{\tiny A}f_3
	\frac{\bp12^{s_{1}+s_{2}+s_{3}}}{\bp23^{s_{1}+s_{3}-s_{2}}\bp31^{s_{2}+s_{3}-s_{1}}}\,
	\partial_{\chi_{1}}^{\,s_{1}}\,\partial_{\chi_{2}}^{\,s_{2}}\,\partial_{\bar\chi_{3}}^{\,s_{3}}	+
	g_\textrm{\tiny H}f_3\frac{\kp12^{s_{1}+s_{2}+s_{3}}}{\kp23^{s_{1}+s_{3}-s_{2}}
	\kp31^{s_{2}+s_{3}-s_{1}}}\,
	\partial_{\bar\chi_{1}}^{\,s_{1}}\,\partial_{\bar\chi_{2}}^{\,s_{2}}\,\partial_{\chi_{3}}^{\,s_{3}}\,.
\end{gathered}
\label{min match}
\ee
where we have made the identifications:
\be
	g_\textrm{\tiny H}\sim g_\textrm{\tiny min}-g_\textrm{\tiny min,PO}\,,
	\qquad
g_\textrm{\tiny A}\sim g_\textrm{\tiny min}+g_\textrm{\tiny min,PO}\,,
\ee
 and have denoted
\be
	f_2=\left(\frac{\kp12\bp12\kp23\bp23}{\kp31\bp31}\right)^{s_2-s_1}\,,\quad
	f_3=\left(\frac{\kp31\bp31\kp23\bp23}{\kp12\bp12}\right)^{s_3-s_1}\,.
	\label{f f}
\ee
We see then that the minimal-coupling vertices potentially give the scattering amplitudes~\eqref{M3g}
in the last three helicity configurations of~\eqref{h1h2h3}, completing the match among cubic
vertices and on-shell amplitudes. However, the helicity configurations
$\left(\pm s_1, \mp s_2, \pm s_3\right)$ and $\left(\pm s_1, \pm s_2, \mp s_3\right)$
are dressed with vanishing factors $f_2$ or $f_3$\,.
In order to get finite amplitudes  in these helicity configurations,
a certain form of non-locality should be allowed in the cubic vertex that would absorb the vanishing factor.
However, such non-locality may well violate other physical requirements, and we shall not pursue this line here. 
Notice though that we obtain in this way
an indirect explanation of why the interactions with $h_1+h_2+h_3=0$ are not allowed, as these can only be
produced in these problematic helicity configurations (recall we assumed $s_1\leq s_2\leq s_3$).

In the special case of $s_1=s_2<s_3$\,,
the factor $f_2$ becomes one, hence no more singular. In fact, this case is where
the expression~\eqref{CPV} involving $V_1$ is no longer valid and it should 
be replaced by~\eqref{PVspecial}. A new calculation shows
a small deviation: 
the holomorphic and anti-holomorphic coupling constants
of  the second and third line of~\eqref{min match}
are now related to $g_{\rm\sst non}$ and $g_{\rm\sst  non,PO}$ as
\be
	g_\textrm{\tiny H}\sim g_\textrm{\tiny min}+g_\textrm{\tiny min,PO}\,,
	\qquad
g_\textrm{\tiny A}\sim g_\textrm{\tiny min}-g_\textrm{\tiny min,PO}\,,
\ee
and the factor $f_3$ differs from that of~\eqref{f f}:
\be
f_3=\frac{(\kp31\bp31)^{s_3-s_1+1}(\kp23\bp23)^{s_3-s_1}}{(\kp12\bp12)^{s_3-s_1+1}}\,,
\ee
but anyway vanishes for conserved momenta. 
The last case left out is when all spins are equal $s_1=s_2=s_3$\,.
In this case, there is no parity-odd minimal-coupling vertex whereas the amplitude
has the two independent holomorphic and anti-holomorphic pieces.

The origin of this disparity may be again related to that of the problematic amplitudes
for the helicity configurations $\left(\pm s_1, \mp s_2, \pm s_3\right)$ and $\left(\pm s_1, \pm s_2, \mp s_3\right)$\,, where
the cubic vertices do provide the same amplitude structures, but they come with a
factor which vanishes upon imposing momentum conservation.
Actually, momentum conservation is what makes a total derivative vanish
from the cubic interaction point of view. In fact, formally, we can 
construct many cubic interactions which are boundary terms. If we calculate
their corresponding amplitudes, we would get formulas 
dressed again with singular factors like $f_2$ and $f_3$\,.
In the case of a parity-odd vertex $s\!-\!s\!-\!s$ with $s$ derivatives,
one can find vertices which are boundary terms and that presumably can reproduce
the missing amplitudes with a singular factor.

It is worth to comment here that in dimensions higher than four, light-cone and covariant classifications of cubic vertices of totally symmetric massless higher-spin fields have a complete agreement~\cite{Metsaev:2005ar,Manvelyan:2010jr}, and the minimal number of derivatives for a cubic vertex with three spins $s_1\leq s_2\leq s_3$ is $s_2+s_3-s_1$, the so-called Metsaev bound.
In four dimensions, which is the lowest dimension for propagating Fronsdal fields, there is a mismatch between the light-cone and covariant classifications of cubic vertices. This is related to the additional structures in the light-cone gauge~\cite{Bengtsson:1986kh}, that violate the Metsaev bound and cannot be uplifted to covariant vertices in a standard fashion.
The amplitudes with problematic helicity configurations $\left(\pm s_1, \mp s_2, \pm s_3\right)$ and $\left(\pm s_1, \pm s_2, \mp s_3\right)$ would correspond to these additional light-cone vertices.
In fact, the light-cone gauge analysis~\cite{Metsaev:1991mt,Metsaev:1991nb} of the quartic interactions in four dimensions shows that for the closure of Poincar\'e algebra on the mass-shell, all the three-point structures \eqref{M3g} are required, including those that do not have covariant counterpart.

As a final remark, we comment on the fact that while~\eqref{M3g} is a non-perturbative result,
the amplitudes produced from cubic vertices are presumably tree level. Actually, it looks reasonable
that the arguments used to constrain the form of the cubic vertex~\eqref{gen cub} can be applied to
the quantum effective action, as they are based just on gauge and Lorentz invariance. In such a case,
the same non-perturbative conclusion is reached via cubic vertices.

\section{Massive Interactions}
\label{sec: massive}

In this section we analyze the match between three-point amplitudes and cubic vertices in the case
where some of the particles/fields have a mass. There are three main cases to be distinguished, namely
when only one, two, or the three particles are massive. These cases are divided into subcases
depending on the relation among the masses if there are several of them. We will start by discussing
the general way to proceed, and then illustrate it with the two simplest
examples: when one particle is massive, and when two particles of the same mass
interact with a massless one.

\subsection{Generalities}

We first quickly review some known facts about three-point amplitudes and cubic vertices
involving massive fields, then show what will be the general strategy to match them.

\subsubsection*{Amplitudes}

On the amplitude side, the classification of three-point amplitudes with the representation discussed
in \ref{massreps} was explicitly considered in \cite{Conde:2016vxs}. We briefly summarize here the most
salient features of the classification.

When only one particle is massive, the functional form of the amplitude is completely fixed up to a
coupling constant, pretty much as it happens in the massless case. Nonetheless, there is a
restriction on the helicities of the massless particles, say particles 1 and 2, depending on the spin $s_3$
of the massive particle. Namely, we must have $|h_1-h_2|\leq s_3$. This constraint can be physically
understood {as arising from the conservation of momentum and angular momentum,
since the process is allowed for real kinematics.}

If two particles, say 1 and 2, are massive we must separate the cases of equal and different masses.
The equal-mass case is similar to the massless one in the sense that the process is kinematically
forbidden for real momenta. The three-point amplitude contains $2\min(s_1,s_2)+1$ ``coupling constants'',
each accompanying a different functional structure. If the masses are different, then the number of
structures depends on the precise relation between $h_3$ and $s_1,s_2$. This case is kinematically
allowed for real momenta, and as in the case of the one massive leg, one gets a restriction
$|h_3|\leq s_1+s_2$ following from the conservation laws of momentum and angular momentum.

When the three particles are massive the functional form of the amplitude is much less constrained, and
the number of possible kinematic structures grows quite large, being bounded by $(2s_i+1)(2s_j+1)$ if
$s_k$ is the biggest spin ($\{i,j,k\}=\{1,2,3\}$).

\subsubsection*{Vertices}

The classification of parity-even cubic vertices with massive fields is done exactly as in
Section~\ref{3ptV}, with just a few differences
(see \textit{e.g.} \cite{Joung:2012rv} for the details). One is that the gauge condition~\eqref{dC} needs
only be imposed when particle $I$ is massless. Another is that the presence of masses
modifies the $\left[Y_I,u_I\cdot\partial_{x_I}\right]$ commutator in~\eqref{commus} as
\be
	\left[Y_I,u_I\cdot\partial_{x_I}\right]=\frac{m_I^2+m_{I+1}^2-m_{I-1}^2}{2}\,.
\ee
The last difference is that the massless Schouten identity $Y_1\,Y_2\,Y_3\,G\approx0$ is modified to
\be
\label{Sch}
	Y_1\,Y_2\,Y_3\,G+\frac12\left(\mu_1\,G_1^2+\mu_2\,G_2^2+\mu_3\,G_3^2\right)+
	\left(\mu_1\,\mu_2+\mu_2\,\mu_3+\mu_3\,\mu_1\right)Z_1\,Z_2\,Z_3\approx0\,,
\ee
where we are denoting $G_I=G-Y_IZ_I$ and $2\mu_I=m_{I+1}^2+m_{I-1}^2-m_I^2$. With all these
ingredients it is simple to work out the form of the vertices for each of the cases specified above.
For the sake of simplicity, we shall not complete the analysis with the parity-odd vertices for massive fields.
Hence, we will not be able to check if opposite-helicity amplitudes can be independently produced from the local vertices.

\subsubsection*{Match}

As done in Section~\ref{match}, we want to extract the three-point amplitudes from the cubic vertices.
For that, we just need to use formula~\eqref{M3C}, which involves the on-shell form of the fields. For
a massless field, this was given in~\eqref{lc fix}-\eqref{spinor}. Let us derive here the analogue 
for an on-shell massive field. We impose the three Fierz conditions
\be
	\left(\Box-m^2\right) \phi_\textrm{\tiny O-S}(x,u)=0\,,
	\qquad 
	\partial_u^2\, \phi_\textrm{\tiny O-S}(x,u)=0\,,
	\qquad \partial_x\cdot\partial_u\, \phi_\textrm{\tiny O-S}(x,u)=0\,.
	\label{MassiveFierz}
\ee
The solution to the first two conditions is simply given by
\ba
\label{mOS}
	\phi_\textrm{\tiny O-S}(x,u)\eq 
	\int\frac{d^4\l\,d^4\tilde \l}{{\rm vol}(U(2))}\,\delta\!\left(\det(\l^I\,\tilde\l_I)+m^2\right)
	\exp\!\left(\frac i2\,x^{a\dot a}\,\lambda^I{}_{a}\,\tilde \lambda_{I\dot a}\right)\times\nn
	&&\quad\times\,\exp\!\left(u_{a\dot a}\,\frac{\partial^2}{\partial \chi_a\,\partial\bar\chi_{\dot a}}\right)
	\tilde\phi_\textrm{\tiny O-S}(\l^I,\tilde\l_I;\chi,\bar\chi)\,\bigg|_{\chi=0=\bar\chi}\,,
\ea
where the division by the volume of $U(2)$ means that
we quotient by the action of $U(2)$ on the space of $\tilde\phi_\textrm{\tiny O-S}$\,.
This requires to fix an irrep of $U(2)=SU(2)\times U(1)$ that the $\tilde\phi_\textrm{\tiny O-S}$ should carry.
Since the massive particles of the three-point amplitudes
considered in \cite{Conde:2016vxs} 
were in their lowest-weight state of $SU(2)$, we also assume here a lowest-weight representation:
\ba
	\label{G-}
	\cK_-\,\tilde\phi_\textrm{\tiny O-S}
	\eq \left(\l^1{}_a\,\frac\partial{\partial \l^2{}_a}-\tilde\l_2{}_{\dot a}\,\frac{
	\partial}{\partial \tilde\l_1{}_{\dot a}}\right) \tilde\phi_\textrm{\tiny O-S}=0\,,\\
	\cK_0\,\tilde\phi_\textrm{\tiny O-S}
	\eq-\frac12 \left(
	\l^1{}_a\,\frac\partial{\partial \l^1{}_a}
	-\l^2{}_a\,\frac\partial{\partial \l^2{}_a}
	-\tilde\l_1{}_{\dot a}\,\frac\partial{\partial \tilde\l_1{}_{\dot a}}
	+\tilde\l_2{}_{\dot a}\,\frac\partial{\partial \tilde\l_2{}_{\dot a}}
	\right) \tilde\phi_\textrm{\tiny O-S}=s\,\tilde\phi_\textrm{\tiny O-S}\,,
\ea
where we are combining the $SU(2)$ generators defined in~\eqref{cKs} as $\cK_-=-\cK^1_2$ and  $\cK_0=\frac12(\cK^2_2-\cK^1_1)$.
By imposing these conditions,
the field, carrying a massive spin $s$ representation,
has the spin angular momentum $-s$ along the space-like direction:
\be
	Q_{a\dot b}=
	\l^1_{a}\,\tilde \l^1_{\dot b}-\l^2_{a}\,\tilde \l^2_{\dot b}\,.
	\label{Q vector}
\ee
See Appendix \ref{app} for the details.
Finally, by imposing the $U(1)$ condition,\footnote{As we discussed below
the equation~\eqref{Vms}, this is just a free choice we have in selecting a certain value for the representation label $r$\,. We notice here a small deviation with respect to \cite{Conde:2016vxs}, where instead of~\eqref{GII}
 the condition $K\,\tilde\phi_\textrm{\tiny O-S}=\cK_0\,\tilde\phi_\textrm{\tiny O-S}$ is imposed as it was
more natural for the reduction of massive amplitudes to massless ones. 
In this paper, we {make the choice~\eqref{GII},}
which leads to a simpler expression for the on-shell fields
in terms of the spinor-helicity variables.}
\be
	\label{GII}
	K\,\tilde\phi_\textrm{\tiny O-S}
	= \left(\l^I{}_a\,\frac\partial{\partial \l^I{}_a}-\tilde\l_I{}_a\,
	\frac\partial{\partial \tilde\l_I{}_a}\right) \tilde\phi_\textrm{\tiny O-S}=0\,,
\ee
the third Fierz (transversality) condition \eqref{MassiveFierz},
\be
	\lambda^I{}_{a}\,\tilde \lambda_{I\dot a}\,\frac{\partial^2}{\partial \chi_a\,\bar\chi_{\dot a}}\,
	\tilde\phi_\textrm{\tiny O-S}(\l^I,\tilde\l_I;\chi,\bar\chi)=0\,,
\ee
can be solved by
\be
\label{mOS-s}
	\tilde\phi_\textrm{\tiny O-S}(\l^I,\tilde\l_I;\chi,\bar\chi)
	=\frac1{s!}\left(\chi_a\,\lambda^{1\,a}\,\bar\chi_{\dot a}\,\tilde\lambda_{2}{}^{\dot a}\right)^s\,
		\tilde\phi^{(s,-s)}_\textrm{\tiny O-S}(\l^I,\tilde\l_I)\,.
\ee
The wave-function $\tilde\phi^{(s,-s)}_\textrm{\tiny O-S}$ carries now the trivial representation
under $U(2)$\,.
{With all the ingredients above, let us now proceed to perform the explicit match between cubic vertices
and massive three-point amplitudes.}

%

\subsection{Two Massless and One Massive}
\label{ssec: 1m}

The generic form of the amplitude for the interaction of two massless particles, with helicities $h_1$ and $h_2$, with a third massive particle with mass $m$ and spin
angular momentum $-s_3$ along the $Q$ direction~\eqref{Q vector}, is given by
\be
\label{M31m}
	M_3 =f_1(m,\kp34)\,\kp12^{-s_3-h_1-h_2} \kp23^{h_1-h_2+s_3}\kp31^{h_2-h_1+s_3}\, ,
\ee
where we {have} parametrized the momenta as $p_1=\lambda^1\tilde\lambda_1$,
$p_2=\lambda^2\tilde\lambda_2$ and $P_3=\lambda^3\tilde\lambda_3+\lambda^4\tilde\lambda_4$.
In \cite{Conde:2016vxs}, the function $f_1$ was fixed to be constant.
However, because of the condition~\eqref{GII}, here $f_1$ will be instead
a homogeneous function of $\kp34$ with weight $-s_3$. This amplitude is only allowed if
\be
\label{h1h2s3}
	|h_1-h_2|\leq s_3\,.
\ee
Let us recover~\eqref{M31m} and~\eqref{h1h2s3} from the cubic vertices.

Assuming without loss of generality that $s_1\leq s_2$, the most general cubic vertex in this case takes the following form (see \cite{Joung:2012rv} for the derivation):
{\be
\label{interaction 1m}
	C^{\sst\rm TT}=
	\sum_{n=\max\{0,s_3-s_2-s_1\}}^{s_3-s_2+s_1}\,\l_n^{(s_1,s_2,s_3)}\,
	H_1^{\frac{s_3+s_2-s_1-n}{2}}\,H_2^{\frac{s_3+s_1-s_2-n}{2}}\,H_3^{\frac{s_1+s_2-s_3+n}{2}}\,Y_3^n\,,
\ee}
where we have introduced the following combinations
\be
	H_1=Y_2\,Y_3-\frac{m^2}{2}\,Z_1\,,\quad
	H_2=Y_3\,Y_1-\frac{m^2}{2}\,Z_2\,,\quad
	H_3=Y_1\,Y_2+\frac{m^2}{2}\,Z_3\,,
\ee
which are useful because the Schouten identity~\eqref{Sch} can be recast in this case as
\be
	H_1\,H_2\,H_3\approx H_3^2\,Y_3^2\,.
\ee
Using this identity, the number of possible vertices for given spins gets reduced to two or one, as
shown in Table~\ref{T1m}.
\begin{table}[t]
\centering
\begin{tabular}{|c||c|c|c|c|c|}
\hline
$\sigma=s_3\!-\! s_1\!-\!s_2$	&	$\ldots,-4,-2$ &	$\ldots,-3,-1$	& 0 & 1 & $2,3,\ldots$		\\
\hline
$n$ &	0 & 1 &	$0$, $2+\s$	&	$1$, $2+\s$	&	$\s$	\\
\hline
\end{tabular}
\caption{Selection among~\eqref{interaction 1m} of the possible cubic vertices for the interaction of one massive and two massless particles in four dimensions. In some cases, two vertices are possible.}
\label{T1m}
\end{table}
These are the vertices that must be now used in~\eqref{M3C}. The action of the operators $H_I$ and $Y_3$ on the on-shell fields casts as
\be
\label{HY1m}
\begin{aligned}
	-4H_1&\leftrightarrow -\frac{m^4\kp31^2}{\kp12^2\kp34}\partial_{\chi_2}+\frac{m^2\kp23^2}{\kp34}\partial_{\bar{\chi}_2}\,, &
	-4H_2&\leftrightarrow -\frac{m^4\kp23^2}{\kp12^2\kp34}\partial_{\chi_1}+\frac{m^2\kp31^2}{\kp34}\partial_{\bar{\chi}_1}\,, \\
	-4H_3&\leftrightarrow \frac{m^2}{\kp12^2}\partial_{\chi_1}\partial_{\chi_2}+\kp12^2\partial_{\bar{\chi}_1}\partial_{\bar{\chi}_2}\,, &
	-2iY_3&\leftrightarrow m^2\frac{\kp31\kp23}{\kp12\kp34}\,.
\end{aligned}
\ee
The factors of $m$ come from the use of momentum conservation, which tells us that a possible
set of kinematically independent spinor products is $\{\kp12,\kp23,\kp31,\kp14,\kp34\}$. The fact that
$\kp14$ does not appear in $H_I,Y_3$ parallels the fact that it neither appears in the
amplitude~\eqref{M31m}. This is ultimately a consequence of the condition~\eqref{G-}.
 
With expressions~\eqref{HY1m} at hand, it is immediate to reproduce the amplitude~\eqref{M31m} from
the vertices in Table~\ref{T1m}. Omitting coupling constants and factors of 2 and $i$, the vertices with $\s\le 1$ give
\begin{equation}
\begin{aligned}
	&\frac{m^{2s_1+2s_2+2s_3}}{\kp34^{s_3}}\kp12^{-s_{1}-s_{2}-s_{3}}\,\kp23^{s_2+s_3-s_{1}}\,\kp31^{s_3+s_1-s_2}\,
	\partial_{\chi_{1}}^{\,s_{1}}\,\partial_{\chi_{2}}^{\,s_{2}}\\
	&+(-1)^{s_2-s_1}\,\frac{m^{2s_3}}{\kp34^{s_3}}\kp12^{s_1+s_2-s_{3}}\,\kp23^{s_1+s_3-s_2}\,\kp31^{s_3+s_2-s_1}\,
	\partial_{\bar\chi_{1}}^{\,s_{1}}\,\partial_{\bar\chi_{2}}^{\,s_{2}}\,,
\end{aligned}
\end{equation}
which correspond to the helicity configurations $(h_1,h_2)=(\pm s_1,\pm s_2)$. Notice that the two different
off-shell vertices with $\s=0,1$ yield the same three-point on-shell amplitudes. Considering now the
vertices with $\sigma\geq2$, we get
\begin{equation}
\begin{aligned}
	&\frac{m^{2s_1+2s_2+2s_3}}{\kp34^{s_3}}\kp12^{-s_{1}-s_{2}-s_{3}}\,\kp23^{s_2+s_3-s_{1}}\,\kp31^{s_3+s_1-s_2}\,
	\partial_{\chi_{1}}^{\,s_{1}}\,\partial_{\chi_{2}}^{\,s_{2}}\\
	&+(-1)^{s_2}\,\frac{m^{2s_1+2s_3}}{\kp34^{s_3}}\kp12^{s_2-s_1-s_{3}}\,\kp23^{s_3-s_2-s_{1}}\,\kp31^{s_1+s_2+s_3}\,
	\partial_{\chi_{1}}^{\,s_{1}}\,\partial_{\bar\chi_{2}}^{\,s_{2}}\\
	&+(-1)^{s_1}\,\frac{m^{2s_2+2s_3}}{\kp34^{s_3}}\kp12^{s_1-s_2-s_3}\,\kp23^{s_1+s_2+s_3}\,\kp31^{s_3-s_1-s_2}\,
	\partial_{\chi_{1}}^{\,s_{1}}\,\partial_{\bar\chi_{2}}^{\,s_{2}}\\
	&+(-1)^{s_2-s_1}\,\frac{m^{2s_3}}{\kp34^{s_3}}\kp12^{s_1+s_2-s_{3}}\,\kp23^{s_1+s_3-s_2}\,\kp31^{s_3+s_2-s_1}\,
	\partial_{\bar\chi_{1}}^{\,s_{1}}\,\partial_{\bar\chi_{2}}^{\,s_{2}}\,,
\end{aligned}
\end{equation}
which also contains the helicity configurations  $(h_1,h_2)=(\pm s_1,\mp s_2)$. 
Hence, interestingly, these
configurations only occur when $s_3\ge s_1+s_2+2$. This is a restriction not
captured by the amplitude, which allows these configurations also when $s_1+s_2=s_3-1,s_3$.
While it would seem natural that the extra vertices in Table~\ref{T1m} that appear when $\sigma=0,1$
would match these configurations, this does not seem to be the case. We can only conjecture that, analogously
to the massless case, the imposition of momentum conservation is preventing us from seeing these
amplitudes from the cubic vertices.

To finish this subsection, let us also derive the condition~\eqref{h1h2s3} from the cubic-vertex analysis.
One just needs to consider two cases. As we saw, the case where $(h_1,h_2)=(\pm s_1,\mp s_2)$ only
happens when $s_3\ge s_1+s_2+2$, automatically implying that $s_3>s_1+s_2=|h_1-h_2|$. In the case
where $(h_1,h_2)=(\pm s_1,\pm s_2)$, we have to check that $s_3\geq s_2-s_1$, which directly follows from
the fact that the exponent of $H_2$ in~\eqref{interaction 1m} should be non-negative.

\subsection{One Massless and Two Equal Massive}
\label{ssec: 2m}

Let us start again by stating the result for the three-point amplitude obtained in \cite{Conde:2016vxs}.
We take particles 1 and 2 to have mass $m$, while particle 3 is massless.
The spin angular {momenta along the} $Q$ directions~\eqref{Q vector} of the
massive particles are fixed to be $-s_1$ and $-s_2$, while the helicity of the third particle, $h_3$, is free.
Without loss of generality, we assume that the spin of the first massive field is not larger than the second one: $s_1\le s_2$\,.
Denoting the momenta as $P_1=\lambda^1\tilde\lambda_1+\lambda^4\tilde\lambda_4$,
$P_2=\lambda^2\tilde\lambda_2+\lambda^5\tilde\lambda_5$ and $p_3=\lambda^3\tilde\lambda_3$,
the amplitude takes the following form:
\be
\label{M32m}
	M_3 =f_2\Big(m,\kp14,\kp25,\frac{\bp45}{\kp12}\Big)\,\kp12^{s_1+s_2+h_3} \kp23^{s_2-s_1-h_3}\kp31^{s_1-s_2-h_3}\, ,
\ee
with the function $f_2$ equal to\footnote{
In \cite{Conde:2016vxs} the combination appearing in $f_2$ was $\left(1+\frac{\kp15\kp24}{m^2}\frac{\bp45}{\kp12}\right)$.
The minus sign here is due to the definition of the angular bracket adopted there, $\kp{I}{J}=\l^{I\, a}\,\l^{J}_{\ a}$, as opposed to $\la I,J\ra=\l^{I}_{\ a}\,\l^{Ja}$ here.
}
\be
\label{f2.2m}
	f_2=\sum_{k=0}^{2s_1}c_k\!\left(\frac{\kp14}{m},\frac{\kp25}{m}\right)\left(1-\frac{\kp14\kp25}{m^2}
	\frac{\bp45}{\kp12}\right)^{s_1+s_2+h_3-k}\,,
\ee
where the free functions $c_k(x,y)$ were fixed to be constants in \cite{Conde:2016vxs}, but here the condition~\eqref{GII} will
give homogeneous functions of weight $-s_1$ and $-s_2$ in $\kp14$ and $\kp25$ respectively.
Notice we have $2s_1+1$ independent kinematic structures in~\eqref{M32m}, a conclusion
we reproduce below via the analysis of cubic vertices.

\smallskip

In this case, the consistent cubic interaction is given in general dimensions by
{\be
	C^{\sst\rm TT}=
	\sum_{n,m}\,\l_{n,m}^{(s_1,s_2,s_3)}\,
	G^n\,Y_1^{s_1-n-m}\,Y_2^{s_2-n-m}\,Y_3^{s_3-n}\,Z_3^m\,.
	\label{interaction 2m}
\ee}
Involving two parameters $n$ and $m$\,, this type of cubic interactions
contains many {vertices with different} number of derivatives. However,
in four dimensions, again thanks to the Schouten identity~\eqref{Sch},
many of them trivialize. We can see that for two equal-mass massive particles
the identity reduces to
\be 
\label{Schouten 2m}
	Y_1\,Y_2\,Y_3\,G\approx -\frac{m^2}2\,(G-Y_3\,Z_3)^2\,.
\ee
Notice that the tensor structures involved in {this identity}
have different number of derivatives.
In order to remove the ambiguities {due to}~\eqref{Schouten 2m},
we take the representative {vertex} having the least derivatives.
In other words, if a certain vertex contains the four-derivative structure
$Y_1\,Y_2\,Y_3\,G$\,, then we replace it by the RHS of~\eqref{Schouten 2m}{,
having} just two derivatives.
Implementing this rule into the general form of interactions~\eqref{interaction 2m},
we are left with three possibilities:
\begin{enumerate}
\item The first case is where the vertices 
do not involve any $G$\,:
\be
	C^{\sst\rm TT}=
	\sum_{m}\,\l_{0,m}^{s_1\!-\!s_2\!-\!s_3}\,
	Y_1^{s_1-m}\,Y_2^{s_2-m}\,Y_3^{s_3}\,Z_3^m\,.
	\label{interaction 2m-1}
\ee
In this case, the number of vertices
are given by the possible values of $m$\,:
there are $s_1+1$ vertices corresponding to
$m=0,1,\ldots,s_1$.
\item
The second case is where the vertices do not involve any $Y_1$\,:
\be
	C^{\sst\rm TT}=
	\sum_{n}\,\l_{n,s_1-n}^{s_1\!-\!s_2\!-\!s_3}\,
	G^{n}\,Y_2^{s_2-s_1}\,Y_3^{s_3-n}\,Z_3^{s_1-n}\,.
	\label{interaction 2m-2}
\ee
The possible vertices are parameterized by $n$\,:
there are $\min\{s_1,s_3\}$ vertices corresponding to
$n=1,2,\ldots,\min\{s_1,s_3\}$.
We drop the possibility of $n=0$ as it overlaps with the case 1.

\item The final case is where 
the vertices do not involve any $Y_3$\,:
\be
	C^{\sst\rm TT}=
	\sum_{m}\,\l_{s_3,m}^{s_1\!-\!s_2\!-\!s_3}\,
	G^{s_3}\,Y_1^{s_1-s_3-m}\,Y_2^{s_2-s_3-m}\,Z_3^{m}\,,
	\label{interaction 2m-3}
\ee
which can occur only when $s_1\ge s_3$\,. There are $s_1-s_3$ possible vertices
corresponding to $m=0,1,\ldots, s_1-s_3-1$.
We drop the possibility of $m=s_1-s_3$ as it overlaps with the case 2.

\end{enumerate}
Irrespective of what $\min\{s_1,s_3\}$ is, we see that there are always $2s_1+1$
possible vertices, coinciding with the number of different structures
in the three-point amplitude. Moreover it is not difficult to see\footnote{
For the comparison, it is better to rewrite~\eqref{f2.2m} as 
$f_2=\xi^{-s_1+s_2+h_3}\sum_{k=0}^{2s_1}c_k\,\xi^{k}$, where what remains inside the sum
is a polynomial in $\xi$, or alternatively in $\frac{\bp45}{\kp12}$, of order $2s_1+1$.
} that these vertices give
rise to~\eqref{M32m}, with the $c_k$ being linear combinations of the
{$\l_{s_3,m}^{(s_1,s_2,s_3)}$}. It is sufficient to use the following expressions for
the operators $Y_I,Z_3,G$,
\be
\label{YZG}
\begin{gathered}
	\begin{aligned}
	-2iY_1&\leftrightarrow\frac{\kp12\kp31}{\kp23}\frac{m^2}{\kp14}\,\xi\,, &
	-2iY_3&\leftrightarrow-\frac{m^2\kp12}{\kp31\kp23}\,\xi\,\partial_{\chi_3}
	+\frac{\kp31\kp23}{\kp12}\,\xi^{-1}\,\partial_{\bar\chi_3}\,,\\
	-2iY_2&\leftrightarrow\frac{\kp12\kp23}{\kp31}\frac{m^2}{\kp25}\,\xi\,,&
	2Z_3&\leftrightarrow\kp12\bp45\,, \\
	\end{aligned}
	\\
	G\leftrightarrow-\frac{m^2\kp12^3}{\kp31\kp23}\,\frac{m^2\,\xi}{\kp14\kp25}\partial_{\chi_3}
	+\kp12\kp31\kp23\frac{\kp14\kp25}{m^2\,\xi}\left(\frac{\bp45}{\kp12}\right)^2\,\partial_{\bar\chi_3}\,,
\end{gathered}
\ee
where we have denoted
\be
	\xi=1-\frac{\kp15\kp24}{m^2}\,\frac{\bp45}{\kp12}\,.
\ee
The expressions in~\eqref{YZG} 
are obtained by acting the operators $Y_I, Z_3$ and $G$
on the on-shell fields (\textit{cf.} equations~\eqref{lc fix}, \eqref{spinor}
for massless fields and~\eqref{mOS}, \eqref{mOS-s} for the massive ones), and using momentum conservation
to express the twenty spinor products that one can build in terms of just eight of them.
The independent set that {we chose is} the following:
$\{\kp12,\kp31,\kp23,\kp14,\kp15,\kp24,\kp25,\bp45\}$. We can notice that, similarly to what happened in Section~\ref{ssec: 1m},
the products $\kp15,\kp24$ do not appear in~\eqref{YZG}, consistently agreeing with the fact
that the lowest-weight amplitude~\eqref{M32m} does not depend on them either.

\section{Conclusion}
\label{sec:conc}

In this paper we have shown 
how the spinor-helicity three-point amplitudes
can be  produced from the local cubic-interaction vertices.
Relating on-shell local fields to wave-functions in the spinor-helicity variables,
we could derive the relations between the \textit{building blocks} of the vertices
such as $Y_I$'s and $G$ to simple rational functions of spinor contractions, $\la I,J\ra$
or $[I,J]$\,. These relations were then used to find the precise dictionary
between the complete cubic vertices and three-point amplitudes.

Our result shows that most of the amplitude structures 
can be reproduced from the cubic interactions.
In particular, the independence between the holomorphic and anti-holomorphic 
amplitudes could be obtained by including both parity-even and -odd vertices, as 
we have checked through the massless cases.
Nonetheless, there remain several amplitudes which do not appear 
from the cubic interactions if we strictly impose the momentum conservation condition.
However, when the latter condition is relaxed at intermediate levels, we could observe
the missing amplitudes do appear but together with some factors 
which vanish when the momentum conservation condition is imposed.
It seems that from the Poincar\'e covariance of the amplitude, 
the nature of locality is not transparent enough and there do exist more structures
than what is allowed by the locality of the Lagrangian,
although presumably boundary terms may produce these extra structures with singular factors.

 \begin{figure}[h]
 \centering\scriptsize
 \begin{tikzpicture}
 \draw [gray, dotted] (0,-1.6) -- (0,1.6);
 \draw [gray, dotted] (-6.5,0) -- (6.5,0);
\draw [blue, thick, fill=blue!15] (6,1.3) rectangle (1,0.65);
\node at (3.5,0.95) {\bf Cubic Vertices  in AdS};
\draw [green, thick, fill=green!15] (-6,1.3) rectangle (-1,0.65);
\node at (-3.5,0.95) {\bf Cubic Vertices in Minkowski};
\draw [blue, thick, fill=blue!15] (6,-1.3) rectangle (1,-0.65);
\node at (3.5,-1) {\bf CFT Three-Point Functions};
\draw [teal, thin, fill=teal!10] (-6.4,-1.6) rectangle (-0.6, -0.35);
\draw [green, thick, fill=green!15] (-6,-1.3) rectangle (-1, -0.65);
\node at (-3.5,-1) {\bf Three-Point Amplitudes};
\draw [ultra thick, orange, <->] (3.5,0.55) -- (3.5,-0.55);
\node at (3.67,0) {\bf Witten Diagrams};
\node at (-3.47,0) {\bf Feynman Diagrams};
\draw [ultra thick, orange, <->] (-3.5,0.55) -- (-3.5,-0.55);
\draw [ultra thick, purple , <-] (-0.85,1) -- (0.85,1);
\node at (0,0.7) {\bf Flat Limit};
\draw [ultra thick, purple, <-] (-0.85,-1) -- (0.85,-1);
\end{tikzpicture}
\caption{Schematic relation among flat/AdS
Local Field Theories and amplitudes/correlators.
The shaded region in the left-down corner
corresponds to the amplitudes which
satisfy the invariance condition but
are not realized by Local Field Theory. }
\end{figure}
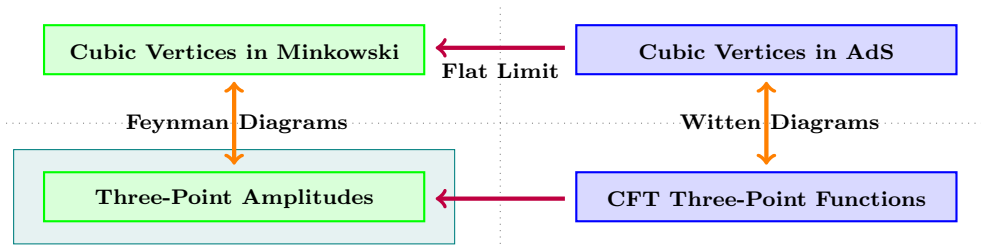

This small discrepancy is somewhat curious when viewing the matching procedure 
as the flat limit of the AdS/CFT duality. 
On the one hand, there are exactly the same number of local cubic interactions in AdS spacetime 
as in flat spacetime \cite{Vasilev:2011xf, Joung:2011ww} (while the classification of deforming and non-deforming vertices differs 
\cite{Boulanger:2012dx,Joung:2013nma}). Actually, the AdS vertices can be obtained from the flat vertices by adding proper 
lower derivative terms required to compensate the non-commutativity of the AdS covariant derivatives.
 See e.g. \cite{Zinoviev:2008ck,Boulanger:2008tg} for  further discussions on the relation between flat and AdS interactions.
On the other hand, independent structures of CFT three-point functions 
can be identified by asking the invariance under the conformal group,
which is isomorphic to the isometry group of AdS \cite{Costa:2011mg,Costa:2011dw}.
In a sense, the CFT three-point functions are the AdS analog of the scattering amplitudes
because
 the way they are determined is the same: by requiring invariance under the isometry group.
It turns out that the number of AdS vertices and the number of CFT three-point functions
exactly match~\cite{Joung:2011ww,Costa:2011mg,Costa:2011dw}: see the recent work~\cite{Sleight:2016dba} where the correspondence between vertex structures of AdS theory and CFT correlator structures of CFT is identified at the three-point level, and used in turn to determine the massless AdS vertices dual to
the three-point functions of the free scalar CFT.
Hence, it seems that there is no discrepancy in the AdS case
like the one present for flat space. It is not clear what makes 
the Poincar\'e invariance --- $\iso(3,1)$ --- differ from the AdS invariance --- $\mathfrak{so}(3,2)$,
but
we presume that it is the algebraic nature:
the latter algebra is simple but the former is not.

\acknowledgments

We thank 
Dario Francia,
Ruslan Metsaev,
Dmitry Ponomarev,
Massimo Taronna
and
Arkady Tseytlin
for useful discussions,
and especially Andrea Marzolla for valuable feedback on the first draft.
The work of EC and EJ was supported by the National Research Foundation of Korea through the grant NRF-2014R1A6A3A04056670.
The work of EJ was supported by the Russian Science Foundation grant 14-42-00047 associated with Lebedev Institute. 
The work of KM was supported by the BK21 Plus Program funded by the Ministry of Education (MOE, Korea) and National Research Foundation of Korea (NRF).
This research was also supported by the Munich Institute for Astro- and Particle Physics (MIAPP) of the DFG cluster of excellence ``Origin and Structure of the Universe''.

\appendix

\section{Identifying massive states}
\label{app}

In this section, we provide an explicit analysis of the
$n=2$ spinor-helicity representation considered in Section \ref{massreps}.
In particular, we will show how a wave-function,
\be
	\la \l,\m\,|\Psi \ra
	=\la \rho_a,\a_a,\b_a,\g_a\,|\Psi\ra\,,
	\label{l m wfn}
\ee
is related to the usual wave-function of a massive particle.
Here, $\l_a$ and $\mu_a$ denote the  two copies of spinors $\l^{I}_a$ ($I=1,2$)\,,
which are parameterized by eight variables $(\rho_a,\a_a,\b_a,\g_a)$ as
\begin{equation}
\begin{aligned}
	\l_1&=\r_1\cos\alpha_1\,e^{i\theta_1}\,, & \l_2&=\r_2\cos\alpha_2\,e^{i\theta_2} \,,\\ 
	\m_1&=\r_1\sin\alpha_1\,e^{i\phi_1}\,, & \m_2&=\r_2\sin\alpha_2\,e^{i\phi_2} \,,
\end{aligned}
\end{equation}
where we have redefined the angular variables as
\begin{equation}
	\a_\pm=\a_1\pm\a_2\,, \quad
	\b_\pm=\th_1-\f_1\pm(\th_2-\f_2)\,, \quad
	\g_\pm=\th_1+\f_1\pm(\th_2+\f_2)\,.
\end{equation}
We take the state vector $|\Psi\ra$ to be the eigenstate of 
momentum $P_{a\dot b}$ and $K$ generators,
$|\Psi\ra=|\,p\otimes r \otimes \psi\,\rangle$\,:
\be
	P_{a\dot b}\,|\,p\otimes r\otimes  \psi\,\rangle
	=p_{a\dot b}\,|\,p\otimes r\otimes \psi\,\rangle\,,
	\qquad
	K\,|\,p\otimes r\otimes  \psi\,\rangle
	=r\,|\,p\otimes r\otimes \psi\,\rangle\,,
	\label{P K cond}
\ee
where $|\psi\ra$ stands for the part of
the state which is not yet determined by
the $P_{a\dot b}$ and $K$ conditions.
For the sake of clarity and simplicity,
we assume henceforth the momentum to be in the rest frame:
\be
	p_{a\dot b}=m\,\delta_{ab}\,.
	\label{P cond}
\ee
This is translated into the following set of equations:
\be
	\r_1=\r_2=\sqrt{m}\,,
	\qquad
	\cos\a_+=0\,,
	\qquad 
	e^{i\,\b_-}=-1\,.
	\label{P var cond}
\ee
After fixing the kinematic condition~\eqref{P var cond}, 
the functional dependence of the wave-function reduces to
just three angular variables, $\a=\a_-\,, \b=\b_+\,, \g=\g_-$, as
\be
	\la \r_a,\a_a,\b_a,\g_a\,|\,p\otimes r \otimes \psi\ra \nn
	 =\delta(\r_1-\sqrt{m})\,\delta(\r_2-\sqrt{m})\,
	\delta(\cos\a_+)\,\delta(e^{i\,\b_-}-1)\,
	e^{i\,r\,\frac{\g_+}4}\,
	\la\a,\b,\g\,|\psi\ra\,.
\ee
The actions of the generators 
$K$ and $\cK^I_J$ 
are realized by the differential operators
$K= -4\,i\,\partial_{\g_+}$, and
\be
	 \cK_0= \frac12\left(\cK_2^2-\cK^1_1\right)=2\,i\,\partial_{\b}\,,\qquad 
	\cK_- =-\cK^1_2=-i\,e^{\frac{i\b}{2}}\left(\partial_\a-2\,i\,\tan\a\,\partial_\b-2\,i\,\sec\a\,\partial_\g\right)\,.
\ee
The latter enjoy the commutation relations of $\mathfrak{su}(2)$:
\begin{equation}
	\left[\cK_+,\cK_-\right]=2\,\cK_0\,,\qquad
	\left[\cK_0,\cK_\pm\right]=\pm\cK_\pm\,.	
	\label{K op}
\end{equation}
where with our conventions $\cK_+=-\cK_-^{\dagger}$.
Now we move to the little group $SO(3)$ operators, which leave
\eqref{P cond} invariant. They are the combinations:
\begin{equation}
\label{J op}
	J_3=L_{12}-\tilde{L}_{\dot 1\dot2}=2\,i\,\partial_{\g}\,,\qquad
	J_-=-L_{11}-\tilde{L}_{\dot 2\dot2}
	=-i\,e^{\frac{i\g}{2}}\left(\partial_\a
	-2\,i\,\tan\a\,\partial_\g-2\,i\,\sec\a\,\partial_\b\right)\,,
\end{equation}
with the commutation relations (again with $J_+=-J_-^{\dagger}$),
\begin{equation}
	\left[J_+,J_-\right]=2\,J_3\,,\qquad
	\left[J_3,J_\pm\right]=\pm\,J_\pm\,.	
\end{equation}
The quadratic Casimir of the above two algebras
coincide:
\ba
	C_2\eq \cK_0{}^2+\frac12\,\{\,\cK_+,\cK_-\}
	=J_3{}^2+\frac12\,\{\,J_+,J_-\}\nn
	\eq-\partial_\a^2+\tan\a\,\partial_\a
	-4\,\sec^2\a\left(\partial_\b^2+\partial_\g^2
	+2\,\sin\a\,\partial_\b\,\partial_g\right)\,.
\ea
Now, we can fix
the state vector $|\psi\ra$ 
to carry a UIR of $\cK_i$ and $J_i$\,.
For instance, we can choose  
$|\psi\ra=|s,\hk,\hj\ra$
to be an eigenstate of $\cK_0$ and $J_0$ 
generators:
\ba
	C_2\,|s,\hk,\hj\ra \eq s(s+1)\,|s,\hk,\hj\ra\,,
	\\
	\cK_0\,|s,\hk,\hj\ra \eq \hk\,|s,\hk,\hj\ra\,,
	\\
	J_3\,|s,\hk,\hj\ra \eq \hj\,|s,\hk,\hj\ra\,.
\ea
From the expressions~\eqref{K op} and~\eqref{J op},
we can conclude that
$\la \a,\b,\g\,|s,\hk,\hj\rangle$ coincides
with the Wigner function,
\be
	\la \a,\b,\g\,|s,\hk,\hj\rangle
	= \la s,\hk|\,R(\a,\b,\g)\,|s,\hj\ra\,,
\ee
where $|s,\hj\ra$ is the eigenstate of $\mathfrak{su}(2)$
and $R(\a,\b,\g)$ is an element of $SU(2)$ with
$(\a,\b,\g)$ related to the Euler angles.
The $\cK_i$ and $J_i$ actions are
realized respectively by the left and right multiplications on the element $R(\a,\b,\g)$\,.

In this paper, we have not diagonalized 
the state vector with respect to $J_3$, but only with respect to $\cK_0$ and $C_2$\,,
hence it remains as a generic linear combination:{
\be
	|\psi^s_{\hk}\ra=\sum_{\hj} c_{\hj}\,|s,\hk,\hj\ra\,.
\ee
}
Defining similarly $|\psi^s\ra=\sum_{\hj}c_{\hj}\,|s,\hj\ra$\,,
we get
\be
	\la \a,\b,\g\,|\psi^s_{\hk}\rangle
	=\la s,\hk|\,R(\a,\b,\g)\,|\psi^s\ra\,.
\ee
Even though the state vector $|\psi^s_{\hk}\ra$ is 
an undetermined one, the 
wave-function  $\la \a,\b,\g\,|\psi^s_{\hk}\rangle$
admits an intuitive interpretation
as the $\vec J\cdot \hat u$ eigenstate with 
eigenvalue $\hk$\,:
\be
	\la \a,\b,\g\,|\, \vec J\cdot \hat u\,|\psi^s_{\hk}\rangle
	=\hk\,	\la \a,\b,\g\,|\psi^s_{\hk}\rangle\,,
\ee
where  the unit vector {$\hat u$ is the 
rotation of $\hat e_3$} by $R(\a,\b,\g)$\,:
\be
	\vec J\cdot \hat u
	=R_J^{-1}(\a,\b,\g)\,J_3\,R_J(\a,\b,\g)\,.
\ee
Relaxing the rest frame condition~\eqref{P cond},
we can also find the `covariant' form of the four-vector
$Q$ --- which reduces to $(0,m\,\hat u)$ in the rest frame --- as
\be
	Q_{a\dot b}=\l_{a}\,\tilde\l_{\dot b}- 
	\m_a\,\tilde\m_{\dot b}\,.
\ee
It is space-like and orthogonal to the momentum vector: 
\be
	Q^2=m^2\,,\qquad 
	P_{a\dot b}\,Q^{a\dot b}=0\,.
\ee
Differently from $P_{a\dot b}$\,,
the vector $Q_{a\dot b}$ does not commute with 
$\cK^I_J$\,,
and most importantly it satisfies
\be 
	Q^{a\dot b}\,W_{a\dot b}
	= -2 m^2\,\cK_0\,,
\ee 
whereas $P^{a\dot b}\,W_{a\dot b}=0$\,.

\bibliography{amplitudesrefs,HSrefs}
\bibliographystyle{JHEP}

\end{document}